\documentclass{crc-book-new}
\usepackage{amsthm}
\usepackage{graphicx}
\usepackage{lscape}
\usepackage{graphicx}
\usepackage{latexsym}
\usepackage{float}
\usepackage{amsthm,,amssymb, color, epsfig, graphics}
\usepackage{xcolor}
\usepackage{graphicx}
\usepackage[intlimits]{amsmath}
\usepackage{latexsym}
\usepackage{amsmath}%[fleqn,intlimits]
\usepackage[all]{xy}

\JNMPnumberwithin{equation}{section}
\newtheorem{proposition}{Proposition}

\theoremstyle{definition}

\def\n{\newline}

\renewcommand{\ba}{\begin{array}}
\renewcommand{\ea}{\end{array}}
\newcommand{\beg}{\begin{eqnarray}}
\newcommand{\eeq}{\end{eqnarray}}
\newcommand{\bg}{\begin{eqnarray*}}

\newcommand{\ed}{\end{eqnarray*}}
\newcommand{\nn}{\nonumber}

\renewcommand{\p}{\partial} 
 
\newcommand{\notlhd}{\lhd\kern-.8em{/}\ } 
\newcommand{\notexist}{\ \exists\kern-.5em{\raise.1em\hbox{/}}\ }

\newcommand{\pde}[2]{\frac{\p #1}{\p #2}} 
 
\newcommand{\pdd}[2]{\frac{\p^2 #1}{\p #2^2}} 
\newcommand{\inp}{{\mbox{\vbox{\hrule width0ex\hbox{\vrule
 height0ex\kern3.8pt
\vbox{\kern2.5pt}\kern3.8pt \vrule height1.6ex}
\hrule width1.6ex}}}}

\usepackage{geometry,multicol}%[centering, margin=20mm]
\newdimen\stockheight
\newdimen\stockwidth
\stockwidth\paperwidth
\usepackage[center,cam]{crop}

\allowdisplaybreaks

\begin{document}

%  Headings
%
\renewcommand{\evenhead}{M Euler and N Euler}
\renewcommand{\oddhead}{{\small CRC Press (2018):} {\small{\it Nonlinear Systems and Their Remarkable Mathematical Structures}}}

%  Titlepage
%
%\thispagestyle{empty}
\thispagestyle{plain}

%\Name{Multipotentialisation of evolution equations: third- order and fifth-order equations}
%\Name{On the multipotentialisation of some symmetry-integrable hierarchies}

\Name{ {\it Chapter C1.} Nonlocal invariance of the multipotentialisations of the Kupershmidt equation and its higher-order hierarchies\,\footnote{Some misprints have been corrected}}

\Author{Marianna Euler and Norbert Euler}

\Address{Division of Mathematics, Department of Engineering Sciences and Mathematics,\\
Lule\aa\ University of Technology, SE-971 87 Lule\aa, Sweden\,\footnote{Current affiliation (since April 2019):
International Society of Nonlinear Mathematical Physics, Auf der Hardt 27,
56130 Bad Ems, Germany \& Centro Internacional de Ciencias, 62210 Cuernavaca, Morelos, Mexico}}
%Auf der Hardt 27,\\
%56130 Bad Ems, Germany
%International Society of Nonlinear Mathematical Physics, 
%Auf der Hardt 27,\\
%56130 Bad Ems, Germany }}
%\\[0.4cm]
%*Current affiliation (since April 2019):\\
%International Society of Nonlinear Mathematical Physics, 
%Auf der Hardt 27,\\
%56130 Bad Ems, Germany \&
%Centro Internacional de Ciencias, Av. Universidad s/n, Colonia Chamilpa,
% 62210 Cuernavaca, Morelos, Mexico\\[0.3cm]
%Corresponding author's email address: Dr.Norbert.Euler@gmail.com}

\noindent
{\bf Publishing details:} This paper is published as {\it Chapter C1} in the book entitled\n {\it ``Nonlinear Systems and Their Remarkable Mathematical Structures''}, N Euler (ed),\n CRC Press, Boca Raton, 317-351, 2018.

%\smallskip

\begin{abstract}
  \noindent
  The term multipotentialisation of evolution equations in $1+1$ dimensions refers to the process of potentialising
  a given evolution equation, followed by at least one further potentialisation of the resulting potential equation. 
  For certain equations this process can be applied several times to result in a finite chain of potential equations, where
  each equation in the chain is a potential equation of the previous equation. By a potentialisation of an equation with dependent variable
  $u$ to an equation with dependent variable $v$, we mean a differential substitution $v_x=\Phi^t$, where $\Phi^t$ is a conserved current of the
  equation in $u$.
  The process of multipotentialisation may lead to interesting nonlocal transformations between the equations. Remarkably, this
  can, in some cases, result
  in nonlocal invariance transformations for the equations, which then serve
  as iteration formulas by which solutions
  can be generated for all the equations in the chain.
  %Not every symmetry-integrable evolution equation can be potentialised and even less equations admit a multipotentialisation.
  %It is therefore of
  %interest to investigate and classify evolution equations by the use of this multipotentialisation process.
  In the current paper we give a comprehensive introduction to this subject and
  report new nonlocal invariance transformations that result from the
  multipotentialisation of the Kupershmidt equation and its higher-order hierarchies. The recursion operators that define the
  hierarchies are given explicitly. 
\end{abstract}

%\noindent
%{\bf Publishing details:} This paper is published in the book {\it Nonlinear Systems and Their Remarkable Mathematical Structures}, N Euler (ed), CRC Press, Boca Raton, 317-351, 2018.

%MSC-class: 37K35, 35C05

%%%%%%%%%%%%%%%%%%%%%%%%

%\renewcommand{\theequation}{\arabic{section}.\arabic{equation}}

%\renewcommand{\theequation}{\arabic{equation}}

%\renewcommand{\theequation}{\arabic{section}.\arabic{equation}}

\section{Introduction: symmetry-integrable equations and \\
  multipotentialisations}

Assume that the following is an $n$th-order symmetry-integrable evolution equation
\begin{gather}
  \label{gen-ee-n}
  E:=u_t-F(x,u,u_x,u_{xx},\ldots,u_{nx})=0.
\end{gather}  
Throughout this paper subscripts $t$ and $x$
    denote partial derivatives. That is, for the dependent variable $u=u(x,t)$, we have
\begin{gather*}
  u_t=\pde{u}{t},\quad u_x=\pde{u}{x},\quad u_{xx}=\pdd{u}{x},\quad  u_{xxx}=\frac{\p^3u}{\p x^3},\quad  u_{qx}=\frac{\p^q u}{\p x^q}
  \quad  \mbox{for}\quad q\geq 4.  
\end{gather*}
For the dependent variable $u_1=u_1(x,t)$ we use the notation $u_{1,x},\ u_{1,xx},\ u_{1,xxx},\  u_{1,qx}$ for $q\geq 4$.
By a symmetry-integrable evolution equation of the form (\ref{gen-ee-n}), we mean an equation that admits a recursion operator $R_u$
that generates an infinite set of commuting
Lie-B\"acklund symmetries
\begin{gather}
    \label{gen-eta_j}
Z_j=\eta_j\pde{\ }{u},\qquad j=1,2,\ldots
\end{gather}
with
\begin{gather}
    \label{gen-Z_j}
\eta_j=R^j_u\,u_t\quad \mbox{or}\quad \eta_k=R_u^k\,u_x\qquad j=1,2,\ldots,\quad k=1,2,\ldots\ .
\end{gather}
Here $\eta_j$ satisfies the symmetry condition
\begin{gather}
  \left.\vphantom{\frac{DA}{DB}}
  L_E[u]\eta_j\right|_{E=0}=0
\end{gather}
for every $j$ for (\ref{gen-ee-n}), where $L_E[u]$ is the linear operator
\begin{gather}
  L_E[u]=\pde{E}{u}+\pde{E}{u_t}D_t+\pde{E}{u_x}D_x+\pde{E}{u_{xx}}D^2_x+\cdots +
  \pde{E}{u_{nx}}D^n_x.
\end{gather}
Here and below, $D_x$ denotes the total $x$-derivative and $D_t$ the total
$t$-derivative. For an extensive list of symmetry-integrable evolution equations we refer to \cite{M}.
%%%
We use integro-differential recursion operators of the form
\begin{gather}
\label{RO-GEN-into}
R_u=\sum_{j=0}^mG_jD_x^j+\sum_{i=1}^3I_i(u_x,u_t)D_x^{-1}\circ \Lambda_i,
\end{gather}
where $G_j$ are functions of $u$ and $x$-derivatives of $u$,
$D_x^{-1}$ is the integral
operator and $\Lambda_i$ are integrating factors of the
equation. 
%%%
The condition for a recursion operator $R_u$ of (\ref{gen-ee-n}) is
\begin{gather}
  \label{Gen-Cond-R}
  \left[L_E[u],\ R_u\right] =D_t R_u,
\end{gather}  
where $[\ ,\ ]$ is the standard Lie bracket $[A,B]:=AB-BA$.
A hierarchy of symmetry-integrable equations then follow for (\ref{gen-ee-n}), namely the hierarchy of evolution equations 
\begin{gather}
  u_{t_j}= R^j_u\,F,\qquad u_{\tau_j}=R^j_u u_x    \qquad j=1,2,\ldots
\end{gather}  
where every equation in the hierarchy admits the infinite set of commuting symmetries (\ref{gen-eta_j}) generated by $R_u^j$.
%\end{document}
Assume that (\ref{gen-ee-n}) admits a conservation law with conserved current $\Phi^t$ and flux $\Phi^x$. That is
\begin{gather}
  \left.\vphantom{\frac{DA}{DB}}
  \left(D_t\Phi^t+D_x\Phi^x\right)\right|_{u_t=F}=0.
\end{gather}
Then $\Phi^t$ must satisfy the relation (see, for example, \cite{Fokas-Fuchssteiner} and \cite{Anco-Bluman})
\begin{gather}
  \label{Phi-t}
  \Lambda=\hat E[u]\,\Phi^t,
\end{gather}
where $\Lambda$ is the corresponding integrating factor (or multiplier) of (\ref{gen-ee-n}) and
$\hat E[u]$ is the Euler operator
\begin{gather}
  \hat E[u]=\pde{\ }{u}-D_x\circ \pde{\ }{u_x}-D_t\circ \pde{\ }{u_t}+D_x^2\circ \pde{\ }{u_{xx}}
  -D_{x}^3\circ \pde{\ }{u_{xxx}}+\cdots\ .
  %\nn\\[0.3cm]
  %\qquad
%  +(-1)^{k}D_x^k\circ \pde{\ }{u_{kx}}.  
\end{gather}
The necessary and sufficient condition on $\Lambda$ is
\begin{gather}
    \hat E[u]\left(\Lambda E\right)=0,
\end{gather}  
or, equivalently, the conditions
\begin{gather}
  \label{alt-cond-Lambda}
  \left.\vphantom{\frac{DA}{DB}}
  L_E^*\Lambda\right|_{E=0}=0\quad
  \mbox{and}\quad
  L_\Lambda E=L^*_\Lambda E,
\end{gather}
where $L^*_E$ is the adjoint operator of $L_E$, namely
\begin{gather}
  \label{L-star}
  L^*_E=\pde{E}{u}-D_x\circ \pde{E}{u_x}-D_t\circ \pde{E}{u_t}
  +D_x^2\circ \pde{E}{u_{xx}}-D_{x}^3\circ \pde{E}{u_{xxx}}+\cdots \nn\\[0.3cm]
  \qquad
  +(-1)^{k}D_x^k\circ \pde{E}{u_{kx}}.  
\end{gather}  
Note that the first condition in (\ref{alt-cond-Lambda}) is the necessary and sufficient condition for an adjoint symmetry $\Lambda$ for
(\ref{gen-ee-n}) (see \cite{Anco-Bluman} for details on adjoint symmetries).

Following the description above, we can calculate
a conservation law of (\ref{gen-ee-n}) by first calculating $\Lambda$ by conditions (\ref{alt-cond-Lambda}) and then
$\Phi^t$ by condition (\ref{Phi-t}). The flux then follows from
\begin{gather}
\label{Phi-x-gen}
  \Phi^x=
\left.\vphantom{\frac{DA}{DB}}
-D_x^{-1}\left(
\vphantom{\frac{DA}{DB}}
D_t\Phi^t\right|_{E=0}\right).
\end{gather}
Using integration by parts, (\ref{Phi-x-gen}) results in the following useful formula for $\Phi^x$:

\begin{proposition}
  \label{Prop-Phi-x-formula}
  Assume that $u_t=F(x,u,u_x,\ldots,u_{nx})$ admits an integrating factor $\Lambda$ and a conserved current
  $\Phi^t=\Phi^t(x,y,u_x,\ldots,u_{mx})$, such that $\Lambda=\hat E[u] \Phi^t$. Then the corresponding flux $\Phi^x$ is given by
\begin{gather}
\label{formula-Phi-x}
  \Phi^x=-D_x^{-1}\left(\Lambda F\right)
  +\sum_{k=1}^m\sum_{j=0}^{m-k} (-1)^k \left(D_x^j F\right)\,D_x^{k-1}\left(
  \pde{\Phi^t}{u_{(j+k)x}}\right).
\end{gather}
\end{proposition}  

\strut\hfill

\noindent
As an example for Proposition \ref{Prop-Phi-x-formula}, consider $\Phi^t=\Phi^t(x,u,u_x,u_{xx},u_{xxx})$.
Then (\ref{Phi-x-gen}) takes the form
\begin{gather}
  \Phi^x=-D_x^{-1}\left(\Lambda F\right)
  -\pde{\Phi^t}{u_x}\,F
  -\pde{\Phi^t}{u_{xx}}\,D_xF
  -\pde{\Phi^t}{u_{xxx}}\, D_x^2 F\\[0.3cm]
  \qquad
  +FD_x\left(\pde{\Phi^t}{u_{xx}}\right)
  -FD_x^2\left(\pde{\Phi^t}{u_{xxx}}\right)
  +(D_x F)\,D_x\left(\pde{\Phi^t}{u_{xxx}}\right).
\end{gather}

\noindent
    {\bf Example:} Consider the Schwarzian Korteweg-de Vries equation (SKdV) 
\begin{gather}
  \label{SKdV}
u_t=u_{xxx}-\frac{3}{2}\frac{u_{xx}^2}{u_x}
\equiv u_x\{u,x\},
\end{gather}
where $\{u,x\}$ it the Schwarzian derivative defined by
\begin{gather*}
  \{u,x\}:=\frac{u_{xxx}}{u_x}-\frac{3}{2}\left(\frac{u_{xx}}{u_x}\right)^2
  \equiv \left(\frac{u_{xx}}{u_x}\right)_x-\frac{1}{2}\left(\frac{u_{xx}}{u_x}\right)^2.
\end{gather*}
%Ref: John Weiss in 1982 (Painlev\'e Property). \n\n
(See \cite{Weiss-1983} where (\ref{SKdV}) was introduced as well as  \cite{Steeb-Euler} and \cite{Conte-1999}
    for more details on its applications in view of the Painlev\'e Test).
Using condition (\ref{alt-cond-Lambda}) we calculate all integrating factors for (\ref{SKdV}) up to order 4 and obtain 
\begin{gather*}
  \Lambda_1=-\frac{2u_{xx}}{u_x^3},\quad \Lambda_2=-\frac{2u^2u_{xx}}{u_x^3}+\frac{4u}{u_x},\quad
  \Lambda_3=\alpha\left(\frac{u_{4x}}{u_x^2}-\frac{4u_{xx}u_{xxx}}{u_x^3}+\frac{3u_{xx}^3}{u_x^4}\right)\\[0.3cm]
  \Lambda_4=\alpha\left(\frac{uu_{xx}}{u_x^3}-\frac{1}{u_x}\right),
\end{gather*}
where $\alpha$ is an arbitrary non-zero constant. Then, using condition (\ref{Phi-t}) and (\ref{formula-Phi-x}), the 
corresponding conserved current and flux for each integrating factor take the following form: 
\begin{gather*}
\Phi_1^t=\frac{1}{u_x},\qquad \Phi_2^t=\frac{u^2}{u_x},\qquad
\Phi_3^t=\frac{\alpha}{2}\left(\frac{u_{xx}}{u_x}\right)^2,\qquad
\Phi_4^t=-\frac{\alpha}{2}\frac{u}{u_x}\\[0.3cm]
\Phi_1^x=\frac{u_{xxx}}{u_x^2}-\frac{1}{2}\frac{u_{xx}^2}{u_x^3},\qquad
\Phi_2^x=\frac{u^2u_{xxx}}{u_x^2}-\frac{1}{2}\frac{u^2u_{xx}^2}{u_x^3}
-4\frac{uu_{xx}}{u_x}+4u_x\\[0.3cm]
\Phi_3^x=\alpha\left(
-\frac{u_{xx}u_{4x}}{u_x^2}
+\frac{1}{2}\frac{u_{xxx}^2}{u_x^2}+\frac{2u_{xx}^2u_{xxx}}{u_x^3}
-\frac{9}{8}\frac{u_{xx}^4}{u_x^4}\right)\\[0.3cm]
\Phi_4^x=
\alpha\left(-\frac{1}{2}\frac{uu_{xxx}}{u_x^2}+\frac{1}{4}\frac{uu_{xx}^2}{u_x^3}+\frac{u_{xx}}{u_x}\right).
\end{gather*}
By the recursion operator Ansatz (\ref{RO-GEN-into}) with $m=2$, condition (\ref{Gen-Cond-R}) results in the following recursion operator for
(\ref{SKdV}) (\cite{Sanders-Wang-2001},  \cite{EE-R-2007}):
%Furthermore, $\Lambda_3$ is needed for the recursion operator of (\ref{SKdV}), which takes the following form:
\begin{gather*}
  R_u=D_x^2-\left(\frac{2u_{xx}}{u_x}\right)D_x+\frac{u_{xxx}}{u_x}-\left(\frac{u_{xx}}{u_x}\right)^2
      -u_xD_x^{-1}\circ \textcolor{black}{\Lambda_3},
\end{gather*}
where $\Lambda_3$ is the integrating factor given above with $\alpha=1$. Now
\begin{gather*}
  \eta_1=R_u \,u_x=u_{xxx}-\frac{3}{2}\frac{u_{xx}^2}{u_x}\\[0.3cm]
  \eta_2=R^2_u\,u_x=u_{5x}-\frac{5u_{xx}u_{4x}}{u_x}-\frac{5}{2}\frac{u_{xxx}^2}{u_x}
  +\frac{25}{2}\frac{u_{xx}^2u_{xxx}}{u_x^2}
  -\frac{45}{8}\frac{u_{xx}^4}{u_x^3},
\end{gather*}
which are two commuting symmetries $Z_1=\eta_1\pde{\ }{u}$ and  $Z_2=\eta_2\pde{\ }{u}$ of (\ref{SKdV}), i.e.
$[Z_1,\ Z_2]=0$. This provides the first two
members of the symmetry-integrable Schwarzian Korteweg-de Vries hierarchy (in terms of the ``time''
variables $t_1$ and $t_2$), namely
\begin{subequations}
\begin{gather*}
  u_{t_1}=u_{xxx}-\frac{3}{2}\frac{u_{xx}^2}{u_x},\ \ 
  u_{t_2}=u_{5x}-\frac{5u_{xx}u_{4x}}{u_x}-\frac{5}{2}\frac{u_{xxx}^2}{u_x}
  +\frac{25}{2}\frac{u_{xx}^2u_{xxx}}{u_x^2}
  -\frac{45}{8}\frac{u_{xx}^4}{u_x^3}.
\end{gather*}
\end{subequations}

\strut\hfill

In the current paper we are interested in the potentialisation of equations. To potentialise equation (\ref{gen-ee-n}) we need to introduce a new variable
$v(x,t)$ such that (\ref{gen-ee-n}) maps to a new equation,
\begin{gather}
  \label{1st-pot-G}
v_t=G(x,v_x,v_{xx},\ldots,v_{\textcolor{black}{n}x}),
\end{gather}
called the potential equation of (\ref{gen-ee-n}). This may be achieved by using an appropriate
conserved current $\Phi_1^t$ and flux $\Phi_1^x$ of (\ref{gen-ee-n}), whereby we let
\begin{gather*}
v_x=\Phi_1^t(x,t,u,u_x,\ldots, u_{qx}),\qquad 
v_t=-\Phi_1^x(x,t,u,u_x,\ldots, u_{qx}).
%\left.\vphantom{\frac{DA}{DB}}
%D_t\Phi^t+D_x\Phi^x\right|_{E=0}=0
\end{gather*}
This is illustrated in Diagram 1.
%\begin{center}
\begin{displaymath}
\qquad\qquad\qquad
  \xymatrix{
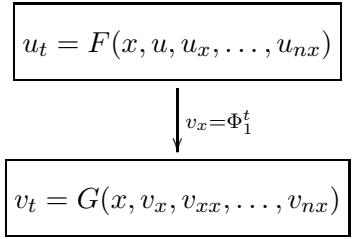
  \mbox{{\bf Diagram 1: Potentialisation of $u_t=F$ }}\\
%A
\boxed{
\vphantom{\frac{DA}{DB}}
u_t=F(x,u,u_x,\ldots,u_{\textcolor{black}{n}x})
}
\ar[d]^{v_x=\Phi_1^t}
\\
%B
\boxed{
\vphantom{\frac{DA}{DB}}
v_t=G(x,v_x,v_{xx},\ldots,v_{\textcolor{black}{n}x})
}
}
\end{displaymath}
%\end{center}

\noindent
%Of course not every equation (\ref{gen-ee-n}) can be potentialised, and this depends on the conservation laws admitted by (\ref{gen-ee-n}).
%In particular,
There is a general relation between the order of the equation (\ref{gen-ee-n}) and the functional dependence of $\Phi^t$ on $u_{qx}$. In particular,
to achieve a potentialisation of an $n$th-order equation (\ref{gen-ee-n}) we need
\begin{gather}
\pde{\Phi^t}{u_{qx}}\neq 0\quad\mbox{with}\quad q\leq n-1,
\end{gather}
where the corresponding integrating factor $\Lambda$ satisfies
\begin{gather}
\pde{\Lambda}{u_{rx}} \neq 0\quad\mbox{with}\quad r\leq n+1\ \ \mbox{(even)} 
\end{gather}
and $r=2q$. The Euler operator $\hat E$ is then of order $q$.

\strut\hfill

We now discuss the multipotentialisation of evolution equations. 
Equation (\ref{gen-ee-n}) can be multipotentialised if (\ref{1st-pot-G}) further admits a potentialisation
\begin{gather}
  w_t=H(x,w_x,\ldots,w_{nx}),
\end{gather}
with
\begin{gather*}
w_x=\Phi_2^t(x,v,v_x,\ldots, v_{qx}),\qquad 
w_t=-\Phi_2^x(x,v,v_x,\ldots, v_{qx}).
\end{gather*}
Here $\Phi_2^t$ is a conserved current and $\Phi_2^x$ a flux for (\ref{1st-pot-G}). This multipotentialisation then results in 
a chain of three equations connected by their potentialisations as illustrated in Diagram 2.

A given equation (\ref{gen-ee-n}) may admit more than one potentialisation, in which case the equation would be related to different
potential equations. Also an equation may potentialise to the same equation by different potentialisations which can lead to interesting invariance transformations. Examples of multipotentialisations and their applications are given in \cite{EE-KN}, \cite{ME-NE-Class-RO-2011},
\cite{EE-converse}, \cite{EE-nonlocal} and \cite{E-E-Reyes}.

%%% ��������������������������

%\begin{center}

  \begin{displaymath}
\qquad\qquad\qquad
    \xymatrix{
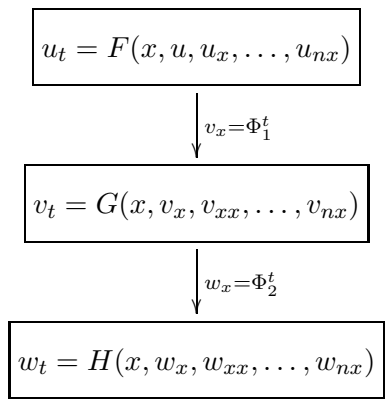
\mbox{{\bf Diagram 2: A multipotentialisation of $u_t=F$}}\\
%A
\boxed{
\vphantom{\frac{DA}{DB}}
u_t=F(x,u,u_x,\ldots,u_{\textcolor{black}{n}x})
}
\ar[d]^{v_x=\Phi_1^t}
\\
%B
\boxed{
\vphantom{\frac{DA}{DB}}
v_t=G(x,v_x,v_{xx},\ldots,v_{\textcolor{black}{n}x})
}
\ar[d]^{w_x=\Phi_2^t}
\\
%B
\boxed{
\vphantom{\frac{DA}{DB}}
w_t=H(x,w_x,w_{xx},\ldots,w_{\textcolor{black}{n}x})
}
}
\end{displaymath}
%\end{center}

%\noindent
%Of course a given equation (\ref{gen-ee-n}) may admit more than one potentialisation so that the equation would be related to different potential equations.
%Also an equation may potentialise to the same equation by different potentialisations which can lead to interesting invariance transformations.

%\strut\hfill

\noindent
{\bf Example:} 
We again consider the
Schwarzian Korteweg-de Vries equation (\ref{SKdV}), namely
\begin{gather*}
%\label{SKdV}
  u_t=u_{xxx}-\frac{3}{2}\frac{u_{xx}^2}{u_x}.
\end{gather*}  
Using the integrating factor
$
\displaystyle{
  \Lambda_1=-\frac{2u_{xx}}{u_x^3}
  }
$
we can potentialise (\ref{SKdV}) with
\begin{gather*}
\displaystyle{
  w_{1,x}=\frac{1}{u_x}
  }
\end{gather*}
to the same equation (\ref{SKdV}), albeit in terms of the variable $w_1$. 
Also, the integrating factor
$
\displaystyle{
  \Lambda_2=-\frac{2u^2u_{xx}}{u_x^3}+\frac{4u}{u_x}
  }
$
leads to
\begin{gather*}
\displaystyle{
  w_{2,x}=\frac{u^2}{u_x}
  }
\end{gather*}
which again satisfies (\ref{SKdV}), now in terms of $w_2$. Hence we have a multipotentialisation of (\ref{SKdV}) to itself. This is illustrated in
Diagram 3. The potentialisations give the relation $w_{2,x}=u^2w_{1,x}$
which leads to a nonlocal invariance transformation $u\mapsto \{w_1,\, w_2\}$ for (\ref{SKdV}), namely
\begin{subequations}
  \begin{gather}
    \label{SKdV-w2}
    w_2=\int \left(\frac{u^2}{u_{x}}\right)\ dx+f_2(t)\\[0.3cm]
    \label{SKdV-w1}
  w_1=\int \left(\frac{1}{u_x}\right)\,dx +f_1(t),
\end{gather}
\end{subequations}
where $f_1$ and $f_2$ have to be determined such that $w_1$ and $w_2$ satisfy (\ref{SKdV}) for any solution $u$ of (\ref{SKdV})
for which $u_x\neq 0$.
%%%%%%%%% MOVE FIGURE

{\small
{
%\begin{center}
\begin{displaymath}
  \xymatrix{
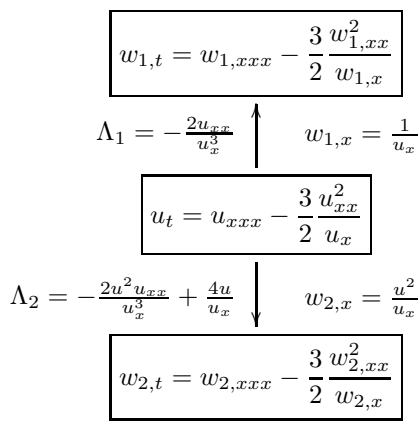
    \mbox{{\bf Diagram 3: Multipotentialisation of SKdV to itself.}}\\
%\mbox{{\ }}\\
%A
%%%5\\
\boxed{
\vphantom{\frac{DA}{DB}}
w_{1,t}=w_{1,xxx}-\frac{3}{2}\frac{w_{1,xx}^2}{w_{1,x}}
}
%%%5\\
\\
\ar[u]^{\ba{l}
\quad { \Lambda_1=
-\frac{2u_{xx}}{u_x^3}  }\ea }_{
  \ba{l}
\quad  w_{1,x}=\frac{1}{u_x}
\ea  \quad}
%%%\\
%%%%%%%%%%%%%%%%%%%\\
%B
\boxed{
\vphantom{\frac{DA}{DB}}
u_t=u_{xxx}-\frac{3}{2}\frac{u_{xx}^2}{u_x}
}
%%%
%%%%%%%%\\
\ar[d]^{\ba{l}
\quad { w_{2,x}=\frac{u^2}{u_x}  }\ea }_{
\ba{l}
\Lambda_2=
-\frac{2u^2u_{xx}}{u_x^3}+\frac{4u}{u_x}\ea }
%%%%%%%\\
\\
%B
\boxed{
\vphantom{\frac{DA}{DB}}
w_{2,t}=w_{2,xxx}-\frac{3}{2}\frac{w_{2,xx}^2}{w_{2,x}}
}
}
\end{displaymath}
%\end{center}
}
}

%%%%%%%%

\noindent
Applying the integrating factor
\begin{gather*}
\Lambda_3=\alpha\left(\frac{u_{4x}}{u_x^2}-\frac{4u_{xx}u_{xxx}}{u_x^3}+\frac{3u_{xx}^3}{u_x^4}\right)
\end{gather*}
we can use the corresponding conserved current $\Phi_3^t$ (given in the previous example) to define the
potential variable $w_3$ as follows:
\begin{gather*}
  w_{3,x}=\frac{\alpha}{2}\frac{u_{xx}^2}{u_x^2}.
\end{gather*}
This leads to
\begin{gather*}
  w_{3,t}=w_{3,xxx}-\frac{3}{4}\frac{w_{3,xx}^2}{w_{3,x}}-\frac{3}{2\alpha}w_{3,x}^2.
\end{gather*}
Also 
\begin{gather*}
\Lambda_4=\alpha\left(\frac{uu_{xx}}{u_x^3}-\frac{1}{u_x}\right)
\end{gather*}
leads to a new potential variable $w_4$, where
\begin{gather*}
w_{4,x}=-\frac{\alpha}{2}\frac{u}{u_x},  
\end{gather*}
which results in the equation
\begin{gather*}
w_{4,t}=w_{4,xxx}-\frac{3}{2}\frac{w_{4,xx}^2}{w_{4,x}}+\frac{3\alpha^2}{8}\frac{1}{w_{4,x}}.  
\end{gather*}
These potentialisations are illustrated in Diagram 4.

%%%%%%%%%% MOVE DIAGRAM 4

{\small
{
%\begin{center}
\begin{displaymath}
  \xymatrix{
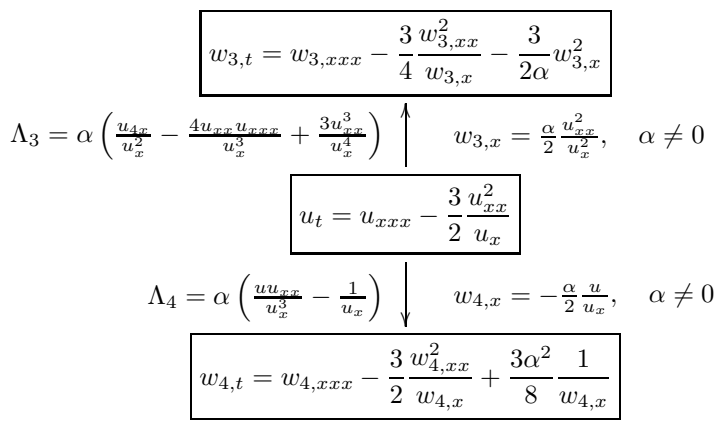
    \mbox{{\bf Diagram 4: Potentialisations of SKdV to two equations.}}\\
%\mbox{{\ }}\\
%A
%%%5\\
\boxed{
\vphantom{\frac{DA}{DB}}
w_{3,t}=w_{3,xxx}-\frac{3}{4}\frac{w_{3,xx}^2}{w_{3,x}} -\frac{3}{2\alpha}w_{3,x}^2
}
%%%5\\
\\
\ar[u]^{\ba{l}
  \quad {
    \Lambda_3=\alpha\left(\frac{u_{4x}}{u_x^2}-\frac{4u_{xx}u_{xxx}}{u_x^3}+\frac{3u_{xx}^3}{u_x^4}\right)
  }\ea }_{
  \ba{l}
\quad  w_{3,x}=\frac{\alpha}{2}\frac{u_{xx}^2}{u_x^2},\quad \alpha\neq 0
\ea  \quad}
%%%\\
%%%%%%%%%%%%%%%%%%%\\
%B
\boxed{
\vphantom{\frac{DA}{DB}}
u_t=u_{xxx}-\frac{3}{2}\frac{u_{xx}^2}{u_x}
}
%%%
%%%%%%%%\\
\ar[d]^{\ba{l}
\quad { w_{4,x}=-\frac{\alpha}{2}\frac{u}{u_x},\quad \alpha\neq 0  }\ea }_{
\ba{l}
\Lambda_4 =\alpha\left(\frac{uu_{xx}}{u_x^3}-\frac{1}{u_x}\right)
\ea }
%%%%%%%\\
\\
%B
\boxed{
\vphantom{\frac{DA}{DB}}
w_{4,t}=w_{4,xxx}-\frac{3}{2}\frac{w_{4,xx}^2}{w_{4,x}}+\frac{3\alpha^2}{8}\frac{1}{w_{4,x}}
}
}
\end{displaymath}
%\end{center}
}
}

%%%%%%%%%%%%%%%%

\noindent
To demonstrate the use of (\ref{SKdV-w2}) -- (\ref{SKdV-w1}) to iterate solutions for the Schwarzian
KdV equation (\ref{SKdV}), we
use the seed solution
\begin{gather*}
  w=e^{x-t/2}+a_1
\end{gather*}
where $a_1$ is an arbitrary constant. By (\ref{SKdV-w1}) we have
\begin{gather*}
  w_1=e^{-x+t/2}+f_1(t),
\end{gather*}  
which satisfies (\ref{SKdV}) iff $f_1(t)=a_2$, where $a_2$ is an arbitrary constant. By (\ref{SKdV-w2}) we then have
\begin{gather*}
  w_2=-a_1^2e^{-x+t/2}+e^{x-t/2}+2a_1x+f_2(t),
\end{gather*}
which satisfies (\ref{SKdV}) iff $f_2(t)=3a_1t+a_3$, where $a_3$ is an arbitrary constant. Thus the given seed solution $w$ leads to two new solutions
\begin{gather*}
  w_1=e^{-x+t/2}+a_2\\[0.3cm]
  w_2=-a_1^2e^{-x+t/2}+e^{x-t/2}+(3t+2x)a_1+a_3.
\end{gather*}
We can now use the above $w_2$ as a new seed solution to generate two more solutions of (\ref{SKdV}) by again applying
(\ref{SKdV-w2}) -- (\ref{SKdV-w1}), etc.

\strut\hfill

We remark that SKdV (\ref{SKdV}) plays a central role in the derivation of iterating solution formulas for the 3rd-order
Krichever-Novikov equation. This has been reported in \cite {EE-KN}.

%%%%%%%%% BEGIN - NEW MATERIAL ABOUT LINEARISABLE 3RD-ORDER EQUATIONS: 29 MAY 2018

\strut\hfill

For some cases an appropriate multipotentialisation can lead to a nonlocal linearisation of the equation.
A remarkable example is the
linearisation of the Calogero-Degasperis-Ibragimov-Shabat (CDIS) equation (see \cite{EE-converse} and \cite{EE-nonlocal})
\begin{gather}
  \label{CDIS}
  u_t=u_{xxx}+3u^2u_{xx}+9uu_x^2+3u^4u_x.
\end{gather}  
Using the integrating factor $\Lambda=2u$, we introduce
\begin{gather}
  \label{1st-pot-v-CDIS}
v_x=u^2,
\end{gather}
by which
(\ref{CDIS}) can be potentialised to
\begin{gather}
\label{1st-pot-CDIS}
  v_t=v_{xxx}-\frac{3}{4}\frac{v_{xx}^2}{v_x}+3v_xv_{xx}+v_x^3.
\end{gather}
Equation (\ref{1st-pot-CDIS}) admits the integrating factor 
\begin{gather*}
\Lambda=
\frac{1}{4}v_x^{-3/2}v_{xx}e^v+\frac{1}{2}v_x^{1/2}e^v,
\end{gather*}
by which we can achieve a potentialisation
of (\ref{1st-pot-CDIS})
with
\begin{gather}
  \label{2nd-pot-w-CDIS}
w_x=v_x^{1/2}e^v
\end{gather}
to the linear equation
\begin{gather}
 \label{2nd-pot-CDIS}
  w_t=w_{xxx}.
\end{gather}
Combining now (\ref{1st-pot-v-CDIS}) and  (\ref{2nd-pot-w-CDIS}), we obtain
\begin{gather}
w_x=ue^{\int u^2 dx}
\end{gather}
or, equivalently,
\begin{gather}
  \label{linearisation-CDIS}
  u_x=\left(\frac{w_{xx}}{w_x}\right)u-u^3,
\end{gather}  
which is a Bernoulli equation that can easily be solved in general to give the linearising transformation
\begin{gather}
  u=w_x\left(2\int w_x^2\,dx+f(t)\right)^{-1/2}
\end{gather}  
of (\ref{CDIS}). Here $f(t)$ must be obtained for every solution $w$ of (\ref{2nd-pot-CDIS}), such that $u$ in
(\ref{linearisation-CDIS}) satisfies (\ref{CDIS}). This multipotentialisation is illustrated in Diagram 5.

\begin{displaymath}
  \xymatrix{
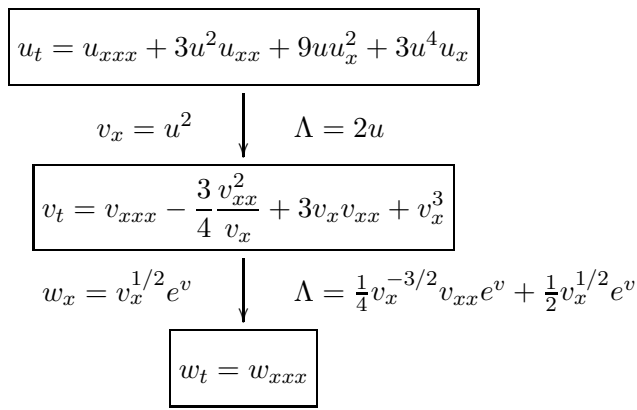
     \mbox{{\bf Diagram 5: Linearisation of the CDIS equation}}\\
%B
\boxed{\vphantom{\frac{DA}{DB}}
u_t=u_{xxx}+3u^2u_{xx}+9uu_x^2+3u^4u_x}%\\
\ar[d]_{
\ba{l}
\textcolor{red}{
{
\textcolor{black}{
v_x=u^2
}}}
\quad\\
%v_t=u_x+u^2
\ea}^{\quad
\ba{l}
\Lambda=2u\ea
}
%\\
\\
%C
\boxed{\vphantom{\frac{DA}{DB}}
v_t=v_{xxx}-\frac{3}{4}\frac{v_{xx}^2}{v_x}+3v_xv_{xx}+v_x^3}%\\
\ar[d]_{
  \ba{l}
\textcolor{red}{
{
\textcolor{black}{
w_x=v_x^{1/2}e^v
}}}
\quad\\
%w_t=v_x e^v
\ea}^{\quad
\ba{l}
\Lambda=\frac{1}{4}v_x^{-3/2}v_{xx}e^v+\frac{1}{2}v_x^{1/2}e^v   \ea
}
%\\
\\
\boxed{\vphantom{\frac{DA}{DB}}
w_t=w_{xxx}
}
}
\end{displaymath}

It is interesting to point out that (\ref{CDIS}) does not admit a recursion operator of the form
(\ref{RO-GEN-into}) with local symmetries $I$ and local integrating factors $\Lambda$ (see \cite{Sanders-Wang-2001}). In \cite{Niklas}
we found that (\ref{CDIS}) admits a recursion operator that generates local symmetries, and therefore a local hierarchy of equations, all of
which are linearisable under the same transformation (\ref{linearisation-CDIS}), namely the operator
\begin{gather*}
%\label{IS-hier}
%u_t=R^n[u]\,u_x, \qquad n=1,2,\ldots\ \nonumber\\[0.5cm]
%\label{recursion_IS}
R_u=D_x^2+2u^2D_x+10uu_x+u^4\notag \\[0.3cm]
\qquad +
2\left(u_{xx}+2u^2u_x
+2ue^{-2\int u^2\,dx}\int \left(e^{2\int u^2\,dx}\,u_x^2\right)\,dx\right)
D_x^{-1}\circ u\notag\\[0.3cm]
%\label{recursion_IS}
\qquad -2ue^{-2\int u^2\,dx}D_x^{-1}\circ
\textcolor{black}{
\left[
\left(u_{xx}+2u^2u_x\right)e^{2\int u^2\,dx}
+2u\int \left(e^{2\int u^2\,dx}\,u_x^2\right)\,dx\right]
}
\nonumber
\end{gather*}
The second member of the higher-order hierarchy is then
\begin{gather*}
%\label{IS-2nd}
u_t=u_{5x}+5u^2u_{4x}+40uu_xu_{xxx}+25uu_{xx}^2+50u_x^2u_{xx}+10u^4u_{xxx}
\notag\\[0.3cm]
\label{IS-2nd*}
\qquad
+120u^3u_xu_{xx}+140u^2u_x^3
+10u^6u_{xx}+70 u^5u_x^2+5u^8u_x.
\end{gather*}

%%%%%%%%% END - NEW MATERIAL ABOUT LINEARISABLE 3RD-ORDER EQUATIONS: 29 MAY 2018

%\strut\hfill

In the current paper we study a chain of symmetry-integrable hierarchies of evolution equations, namely the
chain of the multipotentialisation of the Kupershmidt hierarchy.
This chain of hierarchies is obtained from the Kupershmidt equation,
    \begin{gather*}
      K_t=K_{5x}-5(K_xK_{xxx}+K^2_{xx})-5(K^2K_{xxx}+4KK_xK_{xx}+K_x^3+K^4K_x),
    \end{gather*}
    which is a well-known 5th-order symmetry-integrable evolution equation (\cite{Fordy-Gibbons}, \cite{M}). The paper is organized as
    follows: in Section 1 we perform
    a multipotentialisation of the Kupershmidt equation which results in a chain of five potential equations.
    We also make use of a Miura transformations to map the Kupershmidt equation. In Section 3
    we establish the nonlocal invariance properties for each equation in the chain that was obtained in Section 2.
    In Section 4 we discuss the hierarchies of
    higher-order equations associated with each equation in the chain. We explicitly give the 7th-order equations of the hierarchies.
    In Section 5 we make some concluding remarks and point out some open problems.  The recursion operators that define the hierarchies
    of the Kupershmidt chain are given in Appendix A, and in Appendix B we give an example of a symmetry-integrable equation that
    cannot be potentialised.

\section{The multipotentialisation of the Kupershmidt equation}

In this section we derive a chain of potential Kupershmidt equations by a multipotentialisation of the Kupershmidt equation (see equation
(\ref{kupershmidt})).
We also make use of a result by Fordy and Gibbons
\cite{Fordy-Gibbons} and use Miura transformations to obtain the
Sawada-Kotera equation and an equation that we name the $k$-equation, which is similar to the Sawada-Kotera equation
but with different coefficients.    

\strut\hfill

\noindent
{\bf The Kupershmidt equation} in the dependent variable $K$ is of the form
\begin{gather}
\label{kupershmidt}
\hspace{-0.8cm}\boxed{
  \vphantom{\frac{DA}{DB}}
K_t=K_{5x}
  -5K_xK_{xxx}
  -5K_{xx}^2
  -5K^2K_{xxx}
  -20KK_xK_{xx}
  -5K_x^3
  +5K^4K_x.
  }
\end{gather}  
Equation (\ref{kupershmidt}) is symmetry-integrable and admits a 6th-order recursion operator $R_K$ (\cite{Sanders-Wang-1998},
\cite{ME-NE-Class-RO-2011}) given in Appendix A.
%We us the coefficients of the Kupershmidt equation (\ref{kupershmidt}) according to our paper \cite{ME-NE-Class-RO-2011}, namely eq.(2.9) with $%\alpha=-5$. Note that
%the Kupershmidt equation listed in {Mikailov..} as eq. (4.2.6) and \cite{Sanders-Wang-1998} is eq. (2.9) in \cite{ME-NE-Class-RO-2011} with $\al%pha=5$, for which the recursion operator
%is also given in \cite{Sanders-Wang-1998}.  

By now multipotentialising (\ref{kupershmidt})
we obtain a chain of symmetry-integrable evolution equations of order five, all of which admit recursion operators of order six. Those are listed in
Appendix A.  

\strut\hfill

\noindent
Using the integrating factor
$\Lambda=1$
we introduce the potentialisation
\begin{gather}
  U_x=K.
\end{gather}  
This leads to the {\bf 1st Potential Kupershmidt equation} in the variable $U$ (see Diagram 6), namely
\begin{gather}
  \label{1st-a-pot-kuper}
  \boxed{
  \vphantom{\frac{DA}{DB}}
  U_t=U_{5x}-5U_{xx}U_{xxx}-5U_x^2U_{xxx}-5U_xU_{xx}^2+U_x^5.
  }
\end{gather}

\noindent
We now continue with the 1st Potential Kupershmidt equation (\ref{1st-a-pot-kuper}). With the integrating factor $\Lambda=-e^{2U}$ we introduce
\begin{gather}
u_x=-\frac{1}{2}e^{2U},  
\end{gather}
which leads to the {\bf 2nd Potential Kupershmidt equation}
in the variable $u$ (see Diagram 6), namely
\begin{gather}  
  \label{2nd-pot-kuper}
  \boxed{
  \vphantom{\frac{DA}{DB}}
  u_t=u_{5x}-5\frac{u_{xx}u_{4x}}{u_x}-\frac{15}{4}\frac{u_{xxx}^2}{u_x}
  +\frac{65}{4}\frac{u_{xx}^2u_{xxx}}{u_x^2}
  -\frac{135}{16}\frac{u_{xx}^4}{u_x^3}.
  }
\end{gather}
We remark that the 2nd potential Kupershmidt equation plays a important role in the nonlocal invariance of the Kaup-Kupershmidt equation
(see \cite{Reyes-2005} and  \cite{E-E-Reyes}).

\noindent
Equation (\ref{2nd-pot-kuper}) potentialises in the  
{\bf 3rd Potential Kupershmidt equation}
in the variable $v$ (see Diagram 6)
\begin{gather}
\label{3rd-pot-kuper}   
\boxed{
  \vphantom{\frac{DA}{DB}}
  v_t=v_{5x}-5\frac{v_{xx}v_{4x} }{v_x}+5\frac{v_{xx}^2v_{xxx}}{v_x^2}
  }
\end{gather}
where
\begin{gather}
  v_x=u_x^{-1/2}\quad \mbox{or}\quad v_x=uu_x^{-1/2}.
\end{gather}
These are obtained by the use of the integrating factors
\begin{gather*}
  \Lambda=\frac{3}{4}u_x^{-5/3}u_{xx}\quad \mbox{or}\quad 
  \Lambda=\frac{3}{2}u_x^{-1/2}-\frac{3}{4}uu_x^{-5/2}u_{xx},
\end{gather*}
respectively. Moreover, (\ref{3rd-pot-kuper}) potentialises back into (\ref{2nd-pot-kuper}) by
\begin{gather}
  u_x=v_x^{-2}\quad \mbox{or}\quad u_x=v^4v_x^{-2},
\end{gather}  
which correspond to the integrating factors
\begin{gather*}
  \Lambda=-6v_x^{-4}v_{xx}\quad \mbox{or}\quad \Lambda=12v^3v_x^{-2}-6v^4v_x^{-4}v_{xx},
\end{gather*}
respectively. This fact will be exploited to iterate solutions for these equations (see Propositions \ref{Prop-2nd-pot-invariance}
and Proposition \ref{Prop-3rd-pot-invariance} below).

\strut\hfill

\noindent
Continuing with (\ref{3rd-pot-kuper}) we find that this equation potentialises to the
{\bf 4th Potential Kupershmidt equation}
in the variable $w$ (see Diagram 6)

\strut\hfill

\fbox{
\begin{minipage}{0.93\textwidth}
\begin{gather}
  %\label{4th-pot-kuper}
  w_t=w_{5x}-5\frac{w_{xx} w_{4x}}{w_x}
  -\frac{15}{4}\frac{w_{xxx}^2}{w_x}
  +\frac{65}{4}\frac{w_{xx}^2w_{xxx}}{w_x^2}
  -\frac{135}{16}\frac{w_{xx}^4}{w_x^3}\nn \\[0.3cm]
  \qquad
 \label{4th-pot-kuper}   
+\frac{5\beta}{6}\left(
 \frac{w_{xxx}}{w_x}-\frac{7}{4}\frac{w_{xx}^2}{w_x^2}\right)
 -\frac{5\beta^2}{36}\frac{1}{w_x},\\
 \nn
\end{gather}
\end{minipage}
}

\strut\hfill

\noindent
where
\begin{gather}
  w_x=-\frac{\beta}{6}v^2v_x^{-2}\quad \mbox{with}\quad \Lambda=-\beta vv_x^{-2}+\frac{\beta}{2}v^2v_x^{-4}v_{xx}.
\end{gather}  

%\strut\vfill

%\pagebreak

\begin{center}
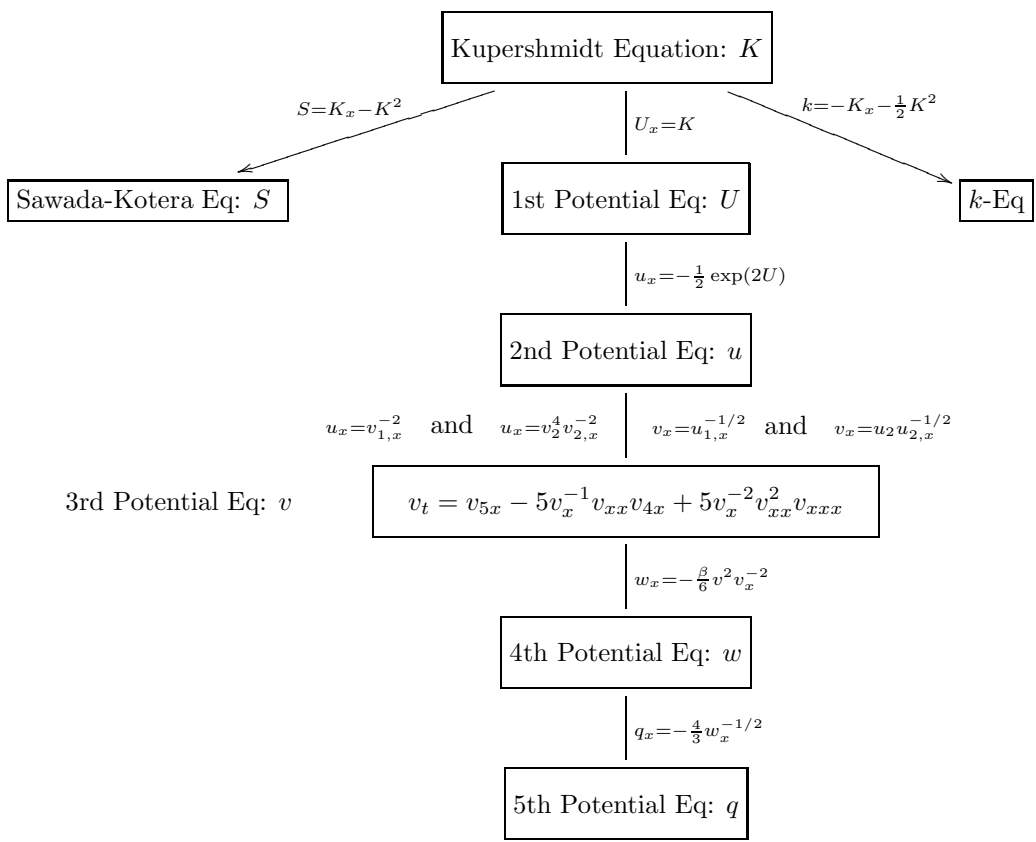
\textcolor{black}{{\bf Diagram 6: A chain of potential Kupershmidt equations}}
\end{center}

%\xymatrix{
%  &Pot. Eq -b\\
%  &\ar@{-}[u]^{a}
%\framebox{Kupershmidt Equation}
%\ar[dr]^{\ \ \ \ \ \ k=-K_x-\frac{1}{2}K^2\ \ \ \ }\ar[dl]_{S=K_x-K^2\ \ \ }\\
%\framebox{Sawada-Kotera Eq}
%&\qquad
%&\framebox{Special Kupershmidt Eq}
%}
%
%\strut\vfill
%
%
%\pagebreak

{\small
\xymatrix{
%%%%
%  & \boxed{
%\vphantom{\frac{DA}{DB}}
%\mbox{1st-(b) Potential Eq: $r$}
%}\\
%  &\ar@{-}[u]_{r_x=\frac{\alpha}{2}K^2}
%
%
  %
&\hspace{-0.5cm}  
  \boxed{
\vphantom{\frac{DA}{DB}}
\mbox{Kupershmidt Equation: $K$}
}
  %\ar@{-}[u]
\ar[dr]^{\ \ \ \ \ \ k=-K_x-\frac{1}{2}K^2\ \ \ \ }\ar[dl]_{S=K_x-K^2\ \ \ }\\
\framebox{Sawada-Kotera Eq: $S$ }
%&\qquad
&\ar@{-}[u]^{\ }_{U_x=K} 
%\boxed{
%\vphantom{\frac{DA}{DB}}
%\mbox{Kupershmidt Eq}}
\boxed{
\vphantom{\frac{DA}{DB}}
\mbox{1st Potential Eq: $U$}
}
&\framebox{$k$-Eq}
& \textcolor{blue}{}  \\
& \ar@{-}[u]_{u_x=-\frac{1}{2}\exp\left(2U\right)}
%\\
%&
%B
\boxed{
\vphantom{\frac{DA}{DB}}
\mbox{2nd Potential Eq: $u$}
}\\
%%% 1
%%% 1
%A 
      {\ \ \ \ \ \ \ \mbox{3rd Potential Eq: $v$}}&\ar@{-}[u]^{u_x=v_{1,x}^{-2}\ \mbox{\textcolor{black}{{\small \ \ and\ \ }}}\ u_x=v_2^4v_{2,x}^{-2}\ \  }_{\ \ v_x=u_{1,x}^{-1/2}
        \mbox{\textcolor{black}{{\small\ \ and\ \ }}}
        \ v_x=u_2u_{2,x}^{-1/2}}
\boxed{
\vphantom{\frac{DA}{DB}}
\ \ \ \displaystyle{v_t=v_{5x} -5v_x^{-1}v_{xx}v_{4x} +5v_x^{-2} v_{xx}^2v_{xxx}}    \ \ \ 
}
& \textcolor{blue}{}\\
%%%%%%%%%%%%%
%A 
&\ar@{-}[u]_{w_{x}=-\frac{\beta}{6}v^2v_x^{-2}}
\boxed{
\vphantom{\frac{DA}{DB}}
\mbox{4th Potential Eq: $w$}
}
&\textcolor{blue}{}   \\
%A 
&\ar@{-}[u]_{q_x=-\frac{4}{3} w_{x}^{-1/2} }
\boxed{
\vphantom{\frac{DA}{DB}}
\mbox{5th Potential Eq: $q$}
}
& \textcolor{blue}{}
%\\
%C
%A
%%%
%% HERE WE CONTROL THE WIDTH OF THE FIRST COLUMN
%\qquad\qquad\qquad&\ar@{-}[u]_{(w_2)_x=-\frac{16}{9}q_x^{-2} }
%%
%\boxed{
%\vphantom{\frac{DA}{DB}}
%\mbox{$W$-Equation}
%}
%\ar[r]^{w_x=\frac{\beta}{6\lambda_0}\exp\left[-2W\right]\quad\ \ }&\ar@{-}[u]_{w_{2,x}=-\frac{16}{9}q_x^{-2} }
%\boxed{
%\vphantom{\frac{DA}{DB}}
%4^{\mbox{th}}\mbox{-Potential Eq $(-)$: $w_2$}
%}
%& \textcolor{blue}{}
%%% 1
%%% 1
%}
}
}

%\begin{gather*}
%  \mbox{\textcolor{blue}{The $W$ equation:}}\\[0.5cm]
%  W_t=W_{5x} +5\left(W_{xx}-W_x^2+\lambda_0 e^{2W}\right)W_{xxx}
%  -5W_xW_{xx}^2\\[0.3cm]
%  \qquad +15\lambda_0W_xW_{xx}e^{2W}+W_x^5+5\lambda_0^2W_x e^{4W}
%\end{gather*}  
%

\strut\hfill

\noindent
Finally we obtain the  {\bf 5th Potential Kupershmidt equation}
in the variable $q$ (see Diagram 6)
\begin{gather}
  \label{5th-pot-kuper}
  \boxed{
  \vphantom{\frac{DA}{DB}}
  q_t=q_{5x}-5\frac{q_{xx}q_{4x}}{q_x}
  +5\frac{q_{xx}^2 q_{xxx}}{q_x^2}
+\frac{15\beta}{32}q_x^2q_{xxx}
-\frac{15\beta}{32}q_xq_{xx}^2
+\left(\frac{3\beta}{32}\right)^2q_x^5
}
\end{gather}
by the potentialisation
\begin{gather}
  q_x=-\frac{4}{3}w_x^{-1/2}\quad \mbox{with}\quad \Lambda=w_x^{-5/2}w_{xx}.
\end{gather}

\strut\hfill

\noindent
Applying the Miura transformation
\begin{gather}
  S=K_x-K^2
\end{gather}
we map the Kupershmidt equation (\ref{kupershmidt}) to the {\bf Sawada-Kotera equation}  \cite{Fordy-Gibbons}
in the variable $S$ (see Diagram 6)
\begin{gather}
  \label{Sawada-Kotera}
\boxed{
  \vphantom{\frac{DA}{DB}}
  S_t=S_{5x}+5SS_{3x} +5S_xS_{xx}+5S^2S_x.
  }
\end{gather}
%Equation (\ref{Sawada-Kotera}) admits the recursion operator $R_S$ of the form (\ref{RO-GEN}) given in Appendix A.
Also, by applying the Miura transformation
\begin{gather}
  k=-K_x-\frac{1}{2}K^2
\end{gather}
the Kupershmidt equation (\ref{kupershmidt}) maps to an equation that we name the {\bf $k$-equation}
(an equation due to Kupershmidt, communicated to A.P. Fordy and J. Gibbons in a private communication \cite{Fordy-Gibbons}) in the
variable $k$ (see Diagram 6)
\begin{gather}
  \label{private-Kuper}
  \boxed{
  \vphantom{\frac{DA}{DB}}
  k_t=k_{5x}+10kk_{xxx}+25k_xk_{xx}+20k^2k_x.
  }
\end{gather}
%Equation (\ref{private-Kuper}) admits the recursion operator $R_k$ of the form (\ref{RO-GEN}) given in Appendix A.

\strut\hfill

\noindent
Next we consider the relations between the 2nd potential Kupershmidt equation (\ref{2nd-pot-kuper}) and the 3rd potential Kupershmidt equation
(\ref{3rd-pot-kuper})
in more detail and derive invariance transformations for those equations.  

\strut\hfill

%\begin{center}
{\bf Diagram 7: Mapping the 2nd potential Kupershmidt equation}
%\end{center}
{\small
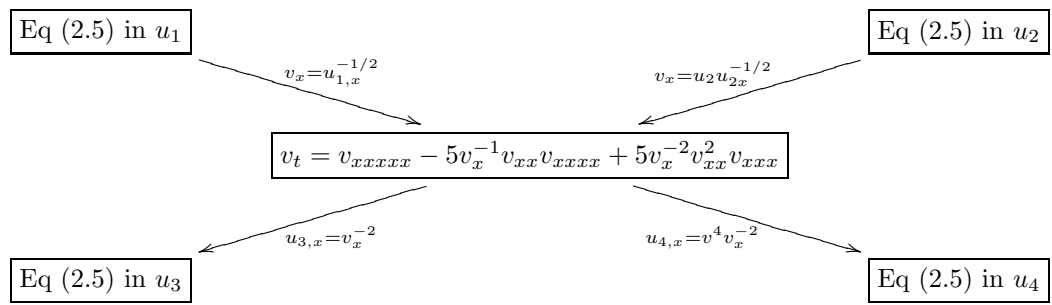
\begin{displaymath}
  \xymatrix{
%&
%2^{{\small \mbox{nd}}}\ \mbox{Potential $\textcolor{blue}{(u)}$ and}N\ 3^{{\small \mbox{rd}}}\ \mbox{Potential $\textcolor{blue}{(v)}$ }\\
%\\
  \boxed{\mbox{Eq (\ref{2nd-pot-kuper}) in } u_1}
\ar[dr]^{v_x=u_{1,x}^{-1/2}}&\qquad &\boxed{\mbox{Eq (\ref{2nd-pot-kuper}) in } u_2}
\ar[dl]_{v_x=u_2u_{2x}^{-1/2}\ \ }\\
&\framebox{$
v_t=v_{xxxxx}-5v_x^{-1} v_{xx}v_{xxxx}+5v_x^{-2} v_{xx}^2v_{xxx}$}
\ar[dr]_{u_{4,x}=v^4v_x^{-2}\ \ \ \ }\ar[dl]^{u_{3,x}=v_x^{-2}}\\
\boxed{\mbox{Eq (\ref{2nd-pot-kuper}) in } u_3}&\qquad & \boxed{\mbox{Eq (\ref{2nd-pot-kuper}) in } u_4}
  }
\end{displaymath}
}
\strut\hfill

\noindent
Diagram 7 depicts the potentialisation of the 2nd potential Kupershmidt equation (\ref{2nd-pot-kuper}) to the 3rd potential Kupershmidt equation
(\ref{3rd-pot-kuper}),
and back again to the 2nd potential Kupershmidt equation; in both cases this is achieved by two different potentialisations as indicated
in Diagram 7.
This leads to

\strut\hfill

\begin{proposition}
  \label{Prop-2nd-pot-invariance}
   \cite{ME-NE-Triangular-2012}
  The 2nd potential Kupershmidt equation (\ref{2nd-pot-kuper}), i.e.
\begin{gather*}
  u_t=u_{5x}-5\frac{u_{xx}u_{4x}}{u_x}-\frac{15}{4}\frac{u_{xxx}^2}{u_x}
  +\frac{65}{4}\frac{u_{xx}^2u_{xxx}}{u_x^2}
  -\frac{135}{16}\frac{u_{xx}^4}{u_x^3},
\end{gather*}  
  is invariant under the transformation $u(x,t)\mapsto \bar u(x,t)$,
  where
  \begin{subequations}
  \begin{gather}
\label{relation-u_1-u_3}
   \bar u=u+c_1\quad\mbox{or}\\[0.3cm]
\label{relation-u2-u3}
   \bar u=-\frac{1}{u}+c_2\quad{or}\\[0.3cm]
\label{relation-u1-u4}
\bar u=\int u_x\left[
\vphantom{\frac{DA}{DB}}
\int u_x^{-1/2}\,dx+f_1(t)\right]^4\,dx+f_2(t).
  \end{gather}
  \end{subequations}
  Here $c_1$ and $c_2$ are arbitrary constants, whereas $f_1$ and $f_2$ have to be determined such that $\bar u$ of (\ref{relation-u1-u4}) satisfies
  (\ref{2nd-pot-kuper}).
\end{proposition}

\noindent
{\bf Proof}: We refer to Diagram 7. Using relations between $u_1$ and $u_3$, we obtain (\ref{relation-u_1-u_3}) with $\bar u=u_3$ and $u=u_1$.
Using relations between $u_2$ and $u_3$, we obtain (\ref{relation-u2-u3}) by eliminating $v_x$ with $\bar u=u_3$
and $u=u_2$. Similarly,
by the given relations between $u_1$ and $u_4$ we eliminate $v$ by differentiation to obtain
\begin{gather}
\label{relation-Bernoulli}
   \bar u_{xx}=\left(\frac{u_{xx}}{u_{x}}\right)\bar u_{x}+4\left(u_{x}^{-1/4}\right)\bar u_{x}^{3/4}.  
  \end{gather}
where $\bar u=u_4$ and $u=u_1$. Equation (\ref{relation-u1-u4}) is a first-order Bernoulli equation of the form
\begin{gather*}
  Y_x=\left(\frac{u_{xx}}{u_{x}}\right)Y+4\left(u_{x}^{-1/4}\right)Y^{3/4},
\end{gather*}
where $Y(x,t)=\bar u_x(x,t)$, which can easily be linearised by the substitution $Z(x,t)=Y^{1/4}(x,t)$. The general solution
is (\ref{relation-u1-u4}).
Note that the condition that results when we make use of the relations between $u_2$ and $u_4$ is the same as the condition which results if we insert
(\ref{relation-u2-u3}) in (\ref{relation-u1-u4}). \qquad $\Box$

\strut\hfill

\noindent
    {\bf Example:} We start with the seed solution
    \begin{gather*}
      u_{1}=-\frac{1}{x}
     \end{gather*} 
 of (\ref{2nd-pot-kuper}). Applying now (\ref{relation-u1-u4}) we obtain
    \begin{gather*}
      \bar u_{1}=
      \frac{1}{112}x^7+\frac{1}{10}f_1(t)x^5+\frac{1}{2}f_1^2(t)x^3+2f_1^3(t)x-f_1^4(t)\left(\frac{1}{x}\right)+f_2(t)
\end{gather*} 
    and insering this $\bar u_{1}$ into (\ref{2nd-pot-kuper}) we obtain
    \begin{gather*}
      f_1(t)=c_{01},\quad f_2(t)=72 c_{01} \,t+c_{02},
     \end{gather*} 
    where $c_{01}$ and $c_{02}$ are arbitrary constants. We could then continue by applying, for example, (\ref{relation-u2-u3}), i.e.
    \begin{gather*}
      \bar u_{2}=-\frac{1}{\bar u_{1}}+c_{03}
    \end{gather*}
    and use the solution $\bar u_{2}$ in (\ref{relation-u1-u4}) to obtain yet another solution $\bar u_{3}$, etc.
    
%
%\strut\hfill
%
%\begin{center}
%{\bf Diagram 3: Invariants of the 3rd potential Kupershmidt equation}
%\end{center}
%{\small
%\begin{displaymath}
%\xymatrix{
%&
%2^{{\small \mbox{nd}}}\ \mbox{Potential $\textcolor{blue}{(u)}$ and}N\ 3^{{\small \mbox{rd}}}\ \mbox{Potential $\textcolor{blue}{(v)}$ }\\
%\\
%  \boxed{\mbox{Eq (\ref{3rd-pot-kuper}) in } v_1}
%\ar[dr]^{u_x=v_{1,x}^{-2}}&\qquad &\boxed{\mbox{Eq (\ref{3rd-pot-kuper}) in } v_2}
%\ar[dl]_{u_x=v_2^4v_{2,x}^{-2}\ \ }\\
%&\framebox{
%$\mbox{2nd Potential Kupershmidt Eq (\ref{2nd-pot-kuper}) in $u$} $}  
%\ar[dr]_{v_{3,x}=uu_x^{-1/2}\ \ \ \ }\ar[dl]^{v_{4,x}=u_x^{-1/2}  }\\
%\boxed{\mbox{Eq (\ref{3rd-pot-kuper}) in } v_4}  &\qquad & \boxed{\mbox{Eq (\ref{3rd-pot-kuper}) in } v_3}
%}
%\end{displaymath}
%}
%
%%%
%%% TRY SOMETHING

\strut\hfill

%\begin{center}
{\bf Diagram 8: Mapping the 3rd potential Kupershmidt equation}
%\end{center}
{\small
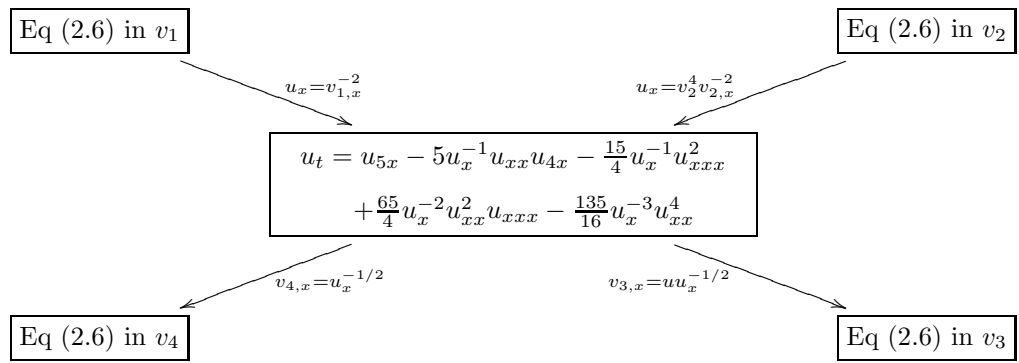
\begin{displaymath}
\xymatrix{
%&
%2^{{\small \mbox{nd}}}\ \mbox{Potential $\textcolor{blue}{(u)}$ and}N\ 3^{{\small \mbox{rd}}}\ \mbox{Potential $\textcolor{blue}{(v)}$ }\\
%\\
  \boxed{\mbox{Eq (\ref{3rd-pot-kuper}) in } v_1}
\ar[dr]^{u_x=v_{1,x}^{-2}}&\qquad &\boxed{\mbox{Eq (\ref{3rd-pot-kuper}) in } v_2}
\ar[dl]_{u_x=v_2^4v_{2,x}^{-2}\ \ }\\
&\framebox{
%$\mbox{2nd Potential Kupershmidt Eq (\ref{2nd-pot-kuper}) in $u$} $}  
  $ \ba{l} {\small u_t=u_{5x}-5u_x^{-1}u_{xx}u_{4x}-\frac{15}{4}u_x^{-1} u_{xxx}^2}\\[0.3cm]
  {\small  \qquad +\frac{65}{4}u_x^{-2}u_{xx}^2u_{xxx}-\frac{135}{16}u_x^{-3}u_{xx}^4 }\ea $
}
%
%u_t=u_{5x}-5\left(\frac{u_{xx}}{u_x}\right)u_{4x}-\frac{15}{4}\left(\frac{1}{u_x}\right)u_{xxx}^2
%+\frac{65}{4}\left(\frac{u_{xx}}{u_x}\right)^2u_{xxx}-\frac{135}{16}\left(\frac{u_{xx}}{u_x}\right)^3u_{xx}
\ar[dr]_{v_{3,x}=uu_x^{-1/2}\ \ \ \ }\ar[dl]^{v_{4,x}=u_x^{-1/2}  }\\
\boxed{\mbox{Eq (\ref{3rd-pot-kuper}) in } v_4}  &\qquad & \boxed{\mbox{Eq (\ref{3rd-pot-kuper}) in } v_3}
}
\end{displaymath}
}

\strut\hfill

%%%&&&&&&&&&&&&&&&&&&&&&&&&&&&&&&&&&&&&&&&&&&&&&&&&&&&&&&&&

\noindent
Diagram 8 depicts the potentialisation of the 3rd potential Kupershmidt equation (\ref{3rd-pot-kuper}) to the 2nd potential Kupershmidt equation (\ref{2nd-pot-kuper}),
and back again to the 3rd potential Kupershmidt equation; in both cases this is achieved by two different potentialisations as indicated in
Diagram 8.
This leads to 

\begin{proposition}
  \label{Prop-3rd-pot-invariance}
  \cite{ME-NE-Triangular-2012}
  The 3rd potential Kupershmidt equation (\ref{3rd-pot-kuper}), i.e.
\begin{gather*}
  v_t=v_{5x}-5\frac{v_{xx}v_{4x} }{v_x}+5\frac{v_{xx}^2v_{xxx}}{v_x^2},
\end{gather*}  
%�����������������������������������
  is invariant under the transformation $v(x,t)\mapsto \bar v(x,t)$,
  where
  \begin{subequations}
  \begin{gather}
    \label{relation-v1-v4}
\bar v=v+c_1, \quad \mbox{or}\\
 \label{relation-v1-v2} 
 \bar v=-\frac{1}{v}+c_2,\quad \mbox{or}\\
 \label{relation v1-v3}
\bar v=v\int \left(\frac{1}{v_x^{2}}\right)\,dx-\int \left(\frac{v}{v_x^2}\right)\,dx+f_1(t)v+f_2(t).
  \end{gather}
  \end{subequations}
  Here $c_1$ and $c_2$ are arbitrary constants, whereas $f_1$ and $f_2$ must be determined such that $\bar v$ of (\ref{relation v1-v3})
  satisfies (\ref{3rd-pot-kuper}).    
\end{proposition}

\strut\hfill

\noindent
{\bf Proof}: We refer to Diagram 8. Using relations between $v_1$ and $v_4$, we obtain (\ref{relation-v1-v4}). Using
the relations between $v_1$ and $v_2$, we obtain (\ref{relation-v1-v2}) by eliminating $u_x$ with $\bar v=v_2$ 
and $v=v_1$. Similarly,
by the given relations between $v_1$ and $v_3$ we eliminate $u$ by differentiation to obtain
\begin{gather}
  \label{Prop-linear-eq}
  \bar v_{xx}=\left(\frac{v_{xx}}{v_{x}}\right)\bar v_{x}+\frac{1}{v_{x}},
\end{gather}
where $\bar v=v_3$ and $v=v_1$. The general solution of the linear equation (\ref{Prop-linear-eq}) is (\ref{relation v1-v3}).
Note that no additional condition results when we make use of the relations between $v_2$ and $v_3$. \qquad $\Box$

\strut\hfill

\noindent
    {\bf Example:} We start with the seed solution
    \begin{gather*}
      v_{1}=\frac{1}{x^2}
    \end{gather*}
    of (\ref{3rd-pot-kuper}). Applying now (\ref{relation v1-v3}) we obtain
    \begin{gather*}
            \bar v_{1}=\frac{1}{70 x^2}\left(
     \vphantom{\frac{DA}{DB}}
     -x^7+70x^2f_2(t)+70f_1(t)\right)
\end{gather*} 
    and insering this $\bar v_{1}$ into (\ref{3rd-pot-kuper}) we obtain
    \begin{gather*}
      f_1(t)=c_{01},\quad f_2(t)=-36\,t+c_{02},
     \end{gather*} 
    where $c_{01}$ and $c_{02}$ are arbitrary constants. We could then continue by applying, for example, (\ref{relation-v1-v2}), i.e.
    \begin{gather*}
      \bar v_{2}=-\frac{1}{\bar v_{1}}+c_{03}
    \end{gather*}
    and use the solution $\bar v_{2}$ in (\ref{relation v1-v3}) to obtain yet another solution $\bar v_{3}$, etc.
    
%\strut\hfill
%
%\noindent
%Using the relations between $v_1$ and $v_3$ given in Diagram 3, we can express solutions $\bar v$ of (\ref{3rd-pot-kuper}) interms of a solution $v$
%of (\ref{3rd-pot-kuper}) and a solution $u$ of (\ref{2nd-pot-kuper}) as follows:
%\begin{subequations}
%\begin{gather}
%   \bar v=uv-\int vv_x^{-2}\,dx +f_2(t)
%   \quad \mbox{and}\quad
%   u=\int v_x^{-2}\,dx+f_1(t),
%\end{gather} 
%\end{subequations}
%where $f_1$ must be determined such that $u$ satisfies (\ref{2nd-pot-kuper}) and $f_2$ such that $\bar v$ satisfies (\ref{3rd-pot-kuper}).

\section{Invariance of the Kupershmidt equation and its chain of potentialisations}

Using Proposition \ref{Prop-2nd-pot-invariance} and Proposition \ref{Prop-3rd-pot-invariance} we obtain nonlocal invariance relations for
all the equations depicted in Diagram 6.

\strut\hfill

\begin{proposition}
\label{Prop-Invariance-Diagram}
\strut\hfill
  
\begin{itemize}
   
\item[a)]
  The Kupershmidt equation (\ref{kupershmidt}) is invariant under $K(x,t)\mapsto \bar K(x,t)$, in which
\begin{gather}
\label{Prop-a}
  \bar K=K+2\left[\ln(v)\right]_x\quad \mbox{and}\quad  K=-\left[\ln(v_x)\right]_x,
\end{gather}
where $v$ is a solution of (\ref{3rd-pot-kuper}) such that $v_x\neq 0$, which can be iterated by the relations in Proposition \ref{Prop-3rd-pot-invariance}.

\item[b)]
  The 1st potential Kupershmidt equation (\ref{1st-a-pot-kuper}) is invariant under $U(x,t)\mapsto \bar U(x,t)$, in which
  \begin{gather}
    \label{Prop-b}
  \bar U=U+2\ln(v)\quad \mbox{and}\quad U=\ln\sqrt{2}-\ln(v_x),
\end{gather}
where $v$ is any solution of (\ref{3rd-pot-kuper}) such that $v_x\neq 0$, which can be iterated by the relations in Proposition \ref{Prop-3rd-pot-invariance}.

\item[c)]
The 2nd potential Kupershmidt equation (\ref{2nd-pot-kuper}) is invariant under $u(x,t)\mapsto \bar u(x,t)$, in which  
\begin{gather}
  \label{Prop-c}
   \bar u=v^4u-4\int uv^3v_x\,dx+f_2(t)
   \quad \mbox{and}\quad
   u=\int \left(\frac{1}{v_x^{2}}\right)\,dx+f_1(t),
\end{gather} 
where $f_1$ and $f_2$ must be determined such that $u$ and $\bar u$ satisfy (\ref{2nd-pot-kuper}).
Here $v$ is any solution of (\ref{3rd-pot-kuper}) such that $v_x\neq 0$, which can be iterated by the relations in Proposition \ref{Prop-3rd-pot-invariance}.

\item[d)]
The 3rd potential Kupershmidt equation (\ref{3rd-pot-kuper}) is invariant under $v(x,t)\mapsto \bar v(x,t)$, in which  
\begin{gather}
  \label{Prop-d}
   \bar v=uv-\int \left(\frac{v}{v_x^{2}}\right)\,dx +f_2(t)
   \quad \mbox{and}\quad
   v=\int u_x^{-1/2}\,dx+f_1(t),
\end{gather} 
where $f_1$ must be determined such that $v$ satisfies (\ref{3rd-pot-kuper}) and $f_2$ such that $\bar v$ satisfies (\ref{3rd-pot-kuper}).
Here $u$ is any solution of (\ref{2nd-pot-kuper}) such that $u_x\neq 0$, which can be iterated by the relations in Proposition \ref{Prop-2nd-pot-invariance}.

\item[e)]
The 4th potential Kupershmidt equation (\ref{4th-pot-kuper}) is invariant under $w(x,t)\mapsto \bar w(x,t)$, in which  
\begin{subequations}
  \begin{gather}
    \label{Prop-e-1}
    \bar w=\left(\frac{\bar v}{uv}\right)^2 w
    -\int w\left[\left(\frac{\bar v}{uv}\right)^2\right]_x\,dx + f_4(t)\\[0.3cm]
    \label{Prop-e-2}
    w=-\frac{\beta}{6}uv^2+\frac{\beta}{3}\int uvv_x\,dx+f_3(t)\\[0.3cm]
    \label{Prop-e-3}
    u=\int \left(\frac{1}{v_x^{2}}\right)\,dx+f_1(t)\\[0.3cm]
    \label{Prop-e-4}
    \mbox{and}\quad
    %\bar v=v\int \left(\frac{1}{v_x^{2}}\right)\,dx -\int \left(\frac{v}{v_x^{2}}\right)\,dx +f_1(t),
    \bar v=\int uu_x^{-1/2}\,dx+f_2(t)
\end{gather}
\end{subequations}
where $f_1$ must be determined such that $u$ satisfies (\ref{2nd-pot-kuper}), $f_2$ such that $\bar v$ satisfies (\ref{3rd-pot-kuper}),
$f_3$ such that
$w$ satisfies (\ref{4th-pot-kuper}), and $f_4$ such that $\bar w$ satisfies (\ref{4th-pot-kuper}). Here $v$ is any solution of (\ref{3rd-pot-kuper})
such that $v_x\neq 0$,
which can be iterated by the relations in Proposition \ref{Prop-3rd-pot-invariance}.

\item[f)]  
The 5th potential Kupershmidt equation (\ref{5th-pot-kuper}) is invariant under $q(x,t)\mapsto \bar q(x,t)$, in which  
\begin{subequations}
  \begin{gather}
    \label{Prop-f-1}
\bar q=q\left(\frac{uv}{\bar v}\right)-\int q \left(\frac{uv}{\bar v}\right)_x\,dx+f_3(t),\quad
q=-\frac{4}{3}\left(\frac{-6}{\beta}\right)^{1/2}\ln (v)\\[0,3cm]
\label{Prop-f-2}
%\bar v=\int uv_x\,dx+f_2(t),\quad \mbox{and}\quad u=\int \left(\frac{1}{v_{x}^{2}}\right)\,dx+f_1(t),
 u=\int \left(\frac{1}{v_x^{2}}\right)\,dx+f_1(t)\quad \mbox{and}\quad
    \bar v=\int uu_x^{-1/2}\,dx+f_2(t)
  \end{gather}
\end{subequations}
where $f_1$ must be determined such that $u$ satisfies (\ref{2nd-pot-kuper}), $f_2$ such that $\bar v$ satisfies (\ref{3rd-pot-kuper}), and $f_3$ such that
$q$ satisfies (\ref{5th-pot-kuper}). Here $\beta<0$ and $v$ is any solution of (\ref{3rd-pot-kuper}) such that $v_x\neq 0$, which
can be iterated by the relations in
Proposition \ref{Prop-3rd-pot-invariance}.

\item[g)] \cite{E-E-Reyes}  
The Sawada-Kotera equation (\ref{Sawada-Kotera})  is invariant under $S(x,t)\mapsto \bar S(x,t)$, in which  
\begin{gather}
  \label{Prop-g}
   \bar S=S+6\left[\ln(v)\right]_{xx}
   \quad \mbox{and}\quad
   S=-\frac{v_{xxx}}{v_x},
\end{gather}
where $v$ is any solution of (\ref{3rd-pot-kuper}) such that $v_x\neq 0$, which can be iterated by the relations in Proposition \ref{Prop-3rd-pot-invariance}.

\item[h)]
    The $k$-equation (\ref{private-Kuper}) is invariant under $k(x,t)\mapsto \bar k(x,t)$, in which 
    \begin{subequations}
      \begin{gather}
        \label{Prop-h-1}
  \bar k=k+\frac{3}{2}\left[\ln(u)\right]_{xx}\\[0.3cm]
\label{Prop-h-2}
  \mbox{and}\quad k=\frac{3}{2}\left[\ln(u)\right]_{xx}
  -\frac{1}{2}\left[\ln(u_x)\right]_{xx}
  -\frac{1}{8}\left[\ln(u_x)\right]_x^2,
\end{gather}
    \end{subequations}
where $u$ is any solution of (\ref{2nd-pot-kuper}) such that $u_x\neq 0$, which can be iterated by the relations in Proposition \ref{Prop-2nd-pot-invariance}.
    
  \end{itemize}
  
\end{proposition}

\strut\hfill

\noindent
    {\bf Proof:} We start with the statement (\ref{Prop-b}) for the 1st potential Kupershmidt equation (\ref{1st-a-pot-kuper}). Following the
    potentialisations as depicted in
Diagram 6, we consider
\begin{gather*}
  \label{Prop-Proof-u1x-u2x}
  u_{1,x}=\frac{1}{2}e^{2U_1},\qquad  u_{2,x}=\frac{1}{2}e^{2U_2},
\end{gather*}
with
\begin{gather*}
  \label{Prop-Proof-u1-u2}
  u_{1,x}=v_x^{-2},\qquad u_{2,x}=v^4v_x^{-2}.
\end{gather*}
This leads to
\begin{gather*}
  \label{Prop-Proof-U1-U2}
  U_2=\ln\left(\sqrt{2}v^2v_x^{-1}\right),\qquad U_1=\ln\left(\sqrt{2}v_x^{-1}\right),
\end{gather*}
so that
\begin{gather*}
  U_2=U_1+\ln(v^2),\qquad   U_1=\ln\sqrt{2}-\ln(v_x).
\end{gather*}  
With $U_2\equiv \bar U$ and $U_1\equiv U$ we obtain (\ref{Prop-b}). Since $U_x=K$, we differentiate (\ref{Prop-b}) to obtain (\ref{Prop-a}).
%By (\ref{Prop-Proof-u1-u2}) we have
Also
\begin{gather*}
  u_{2,x}=u_{1,x}v^4,\qquad u_{1,x}=v_x^{-2}
\end{gather*}
which becomes (\ref{Prop-c}) after integration. Integrating the relations
\begin{gather*}
  v_{1,x}=u_x^{-1/2},\qquad v_{3,x}=uu_x^{-1/2}
\end{gather*}
we obtain
\begin{gather*}
  v_1=\int u_x^{-1/2}\,dx+f_1(t)\\[0.3cm]
  v_3=\int u v_{1,x}\,dx+f_2(t)=\int uv_1\,dx-\int v v_x^{-2}\,dx+f_2(t).
\end{gather*}
With $v_3\equiv \bar v$ and $v_1\equiv v$ we obtain (\ref{Prop-d}).
For the 4th potential Kupershmidt equation (\ref{4th-pot-kuper}) we consider
\begin{gather*}
  w_{1,x}=-\frac{\beta}{6}v_1^2v_{1,x}^{-2},\qquad
  w_{2,x}=-\frac{\beta}{6}v_3^2v_{3,x}^{-2}\\[0.3cm]
  v_{1,x}=u_x^{-1/2},\qquad v_{3,x}=uu_x^{-1/2},
\end{gather*}  
which leads to
\begin{gather*}
  w_{2,x}=w_{1,x}v_1^{-2}v_3^2u^{-2},\qquad
  w_{1,x}=-\frac{\beta}{6}v_1^2u_x. 
\end{gather*}
Integrating the above expressions for $w_{2,x}$ and $w_{1,x}$ leads to (\ref{Prop-e-1}) -- (\ref{Prop-e-4}), where $w_2\equiv \bar w$, $w_1\equiv w$, $v_3\equiv \bar v$ and $v_1\equiv v$.
For the 5th potential Kupershmidt equation (\ref{5th-pot-kuper}) we consider
\begin{gather*}
  q_{1,x}=-\frac{4}{3}w_{1,x}^{-1/2},\qquad q_{2,x}=-\frac{4}{3}w_{2,x}^{-1/2}\\
  w_{1,x}=-\frac{\beta}{6}v_1^2v_{1,x}^{-2},\qquad w_{2,x}=-\frac{\beta}{6}v_3^2v_{3,x}^{-2}\\
  v_{1,x}=u_x^{-1/2},\qquad\quad v_{3,x}=uu_x^{-1/2}.
 \end{gather*}
Combining the above relations, we obtain
\begin{gather*}
  q_{1,x}=A v_1^{-1}u_x^{-1/2},\qquad q_{2,x}=Av_3^{-1}uu_x^{-1/2},\qquad
  A=-\frac{4}{3}\left(-\frac{\beta}{6}\right)^{-1/2},
\end{gather*}  
so that
\begin{gather*}
  q_{2,x}=uv_1v_3^{-1}q_{1,x}\quad \mbox{where}\quad q_{1,x}=Av_1^{-1}v_{1,x}.
\end{gather*}
Integrating the above expressions for $q_{2,x}$ and $q_{1,x}$, we obtain (\ref{Prop-f-1}) -- (\ref{Prop-f-2}), where $q_2\equiv \bar q$, $q_1\equiv q$, $v_3\equiv \bar v$ and $v_1\equiv v$. For the Sawada-Kotera equation
(\ref{Sawada-Kotera}) we consider
\begin{gather*}
  S_1=K_{1,x}-K_1^2,\qquad S_2=K_{2,x}-K_2^2
\end{gather*}
with
\begin{gather*}
  K_2=K_1+6\left[\ln(v)\right]_x,\qquad K_1=-\left[\ln(v_x)\right]_x.
\end{gather*}
This leads to
\begin{gather*}
  S_2=S_1+6\left[\ln(v)\right]_{xx},\qquad S_1=-\frac{v_{xxx}}{v_x},
\end{gather*}  
which is (\ref{Prop-g}) where $S_2\equiv \bar S$ and $S_1\equiv S$. Finally, for the $k$-equation (\ref{private-Kuper})
we have
\begin{gather*}
  k=-K_x-\frac{1}{2}K^2,\qquad
  K=-\left[\ln(v_x)\right]_x.
\end{gather*}
This leads to
\begin{gather*}
  k=\frac{v_{xxx}}{v_x}-\frac{3}{2}\left(\frac{v_{xx}}{v_x}\right)^2\equiv \{v,\,x\},
\end{gather*}
where $\{v,\,x\}$ is the Schwarzian derivative. We now consider
\begin{gather*}
  k_1=\{v_1,\,x\},\qquad
  k_2=\{v_2,\,x\}\\
  v_{1,x}=uu_x^{-1/2},\qquad v_{2,x}=u_x^{-1/2},
\end{gather*}
which leads to (\ref{Prop-h-1}) -- (\ref{Prop-h-2}), where $k_2\equiv \bar k$ and $k_1\equiv k$.\qquad $\Box$

%%#######################################

\strut\hfill

\noindent
    {\bf Example:}
    We apply Proposition \ref{Prop-Invariance-Diagram} e) for the 4th potential Kupershmidt equation (\ref{4th-pot-kuper}). We consider the seed
    solution
    \begin{gather*}
      v=\frac{\sqrt{5}x^4}{20}+\frac{36\sqrt{5}t}{x}
    \end{gather*}
    for the 3rd potential Kupershmidt equation (\ref{3rd-pot-kuper}). By requiring that 
    \begin{gather*}
      u=\int v_{x}^{-2}\,dx+f_1(t)
    \end{gather*}
    satisfies (\ref{2nd-pot-kuper}) we obtain $f_1(t)=c_1$, where $c_1$ is an arbitrary constant. Then
    \begin{gather*}
      u=-\frac{1}{x^5-180 t}+c_1.
    \end{gather*}
    We then calculate $\bar v$ by the formula
    \begin{gather*}
      \bar v=\int uu_x^{-1/2}\,dx+f_2(t)
    \end{gather*}
    and, by requiring that $\bar v$ satisfies (\ref{3rd-pot-kuper}) we obtain $f_2(t)=c_2$, where $c_2$ is an arbitrary constant. Then
    \begin{gather*}
    \bar v=\frac{-\sqrt{5}(c_1x^5+720c_1t+4)}{20x}+c_2.  
    \end{gather*}
    A solution for (\ref{4th-pot-kuper}) is now obtained by the formula
    \begin{gather*}
      w=-\frac{\beta}{6}v^2u+\frac{\beta}{6}\int uvv_{x}\,dx+f_3(t),
    \end{gather*}
    from which it follows that $f_3(t)=c_3$, where $c_3$ is an arbitrary constant. This gives
    \begin{gather*}
      w=\frac{\beta x^8-288c_3 x^5-2880\beta t x^3+51840 c_3t}{-288x^5+51840 t}.
     \end{gather*} 
    One more solution for (\ref{4th-pot-kuper}) then follows by applying the formula
    \begin{gather*}
      \bar w=\left(\frac{\bar v}{uv}\right)^2w-\int w\left(\frac{\bar v^2}{u^2v^2}\right)_x\,dx+f_4(t).
    \end{gather*}
    Again $f_4(t)=c_4$, where $c_4$ is an arbitrary constant. This solution for (\ref{4th-pot-kuper}) then takes the following form:
    \begin{gather*}
      \bar w=\left(
      \beta c_1^2x^8-288c_1^2c_5x^5
      +24\sqrt{5}\beta c_1c_2x^4
      -16\beta c_1x^3
      -2880\beta c_1^2x^3t\right.\\[0.3cm]
      \qquad\left.\vphantom{\sqrt{X^8}}
      +51840 c_1^2c_5 t
      -48\beta c_2^2\right)
      \left( \vphantom{\sqrt{X^8}}
      -288c_1^2x^5+51840c_1^2 t+288c_1\right)^{-1},
    \end{gather*}
where we have set $c_3+c_4=c_5$. 
      
\section{The hierarchies}

All equations in Diagram 6 admit
six-order integro-differential recursion operators of the form
\begin{gather*}
%\label{RO-GEN}
R_u=\sum_{j=0}^6G_jD_x^j+\sum_{i=1}^3I_i(u_x,u_t)D_x^{-1}\circ \Lambda_i,
\end{gather*}
where $G_j$ are functions of $u$ and $x$-derivatives of $u$. $\Lambda_i$ are integrating factors of the equations.

Acting a recursion operator $R_u$ of a symmetry-integrable evolution equation $u_t=F(u,u_x,\ldots,u_{5x})$ on
its $t$-translation symmetry $\displaystyle{u_t\pde{\ }{u}}$ or its $x$-translation symmetry
$\displaystyle{u_x\pde{\ }{u}}$,
%
%i.e.
%\begin{gather}
%  u_{t_j}=R_u^j\,u_t\qquad j=1,2,\ldots,
%\end{gather}
results in a hierarchy of higher-order commuting symmetries and these flows define a higher-order hierarchy of symmetry-integrable
evolution equations. The equations of Diagram 6 result in two symmetry-integrable hierarchies for each equation.
For example, for the Kupershmidt equation (\ref{kupershmidt}) we obtain the following two hierarchies of equations 
%can now be introduced for both Kupershmidt hierarchies, i.e.
\begin{subequations}
\begin{gather*}
  K_{t_j}=R^j_K\,K_t\quad \mbox{is a hierarchy of order $6j+5$, and}\\
  K_{\tau_j}=R^j_K\,K_x\quad \mbox{is a hierarchy of order $6j+1$},
\end{gather*}
\end{subequations}
where $j=1,2,\ldots$ and 
\begin{gather*}
K_t=K_{5x}
  -5K_xK_{xxx}
  -5K_{xx}^2
  -5K^2K_{xxx}
  -20KK_xK_{xx}
  -5K_x^3
  +5K^4K_x.
\end{gather*}
The recursion operator $R_K$ for (\ref{kupershmidt}) is given in Appendix A.
%
%symmetry-integrable evolution equations.
%
%of order $6j+5$.
%Moreover, when acting $R_u$ on the $x$-translation symmetry $u_x$ of $u_t=F(u,u_x,\ldots,u_{5x})$, i.e.
%\begin{gather}
%  u_{\tau_j}=R_u^j\, u_x
%\end{gather}
%we obtain a hierarchy of higher-order symmetry-integrable evolution equations of order $6j+1$.
This is the case for all
the equations in Diagram 6, i.e. the Kupershmidt equation, its 5 potential equations, the Sawada-Kotera equation, and the $k$-equation.
The recursion operators of all these equations are listed explicitly in Appendix A.

The same potential variables $U$, $u$, $v$, $w$, and $q$ that were introduced in Section 2 for the Kupershmidt equation (\ref{kupershmidt}),
illustrated in Diagram 6, can now be introduced for all the corresponding hierarchies. This is illustrated in Diagram 10. In Diagram 9 we illustrate the
Sawada-Kotera hierarchy and the $k$-equation hierarchy related to the Kupershmidt hierarchy by Miura transformations.  
%\begin{subequations}
%\begin{gather*}
%  K_{t_j}=R^j_K\,K_t\qquad \mbox{(hierarchy of order $6j+5$)}\\
%  K_{\tau_j}=R^j_K\,K_x\qquad \mbox{(hierarchy of order $6j+1$)},
%\end{gather*}
%\end{subequations}
%where $j=1,2,\ldots$ and 
%The recursion operator $R_K$ for (\ref{kupershmidt}) is given in Appendix A. This chain of hierarchies for the multipotentialisations of the
%Kupershmidt equation is illustrated in Diagram 4.
The flows corresponding to the different ``times'', $t_j$ and $\tau_j$, all commute for all equations in Diagram 9 and Diagram 10.  That is
\begin{gather*}
  [Z_t,\ Z_{t_j}]=0,\quad
  [Z_t,\ Z_{\tau_j}]=0,\quad
  [Z_{t_i},\ Z_{t_j}]=0\\[0.3cm]
  [Z_{\tau_i},\ Z_{\tau_j}]=0,\quad
  [Z_{t_i},\ Z_{\tau_j}]=0,
\end{gather*}
where, for the equation $u_t=F(u,u_x,u_{xx},\ldots,u_{5x})$, we have
\begin{gather*}
  Z_t=u_t\pde{\ }{u}\equiv F\pde{\ }{u},\quad
  Z_{t_j}=\left(R^j_u\,u_t\right)\pde{\ }{u},\quad
  Z_{\tau_j}=\left(R^j_u\,u_x\right)\pde{\ }{u}.
\end{gather*}

The same invariance transformations as given in Proposition \ref{Prop-2nd-pot-invariance}, Proposition \ref{Prop-3rd-pot-invariance} and
Proposition \ref{Prop-Invariance-Diagram} now also hold for the
respective higher-order hierarchies of equations illustrated in Diagram 9 and Diagram 10.

{\small
{
%\begin{center}
\begin{displaymath}
\qquad\ \ 
  \xymatrix{
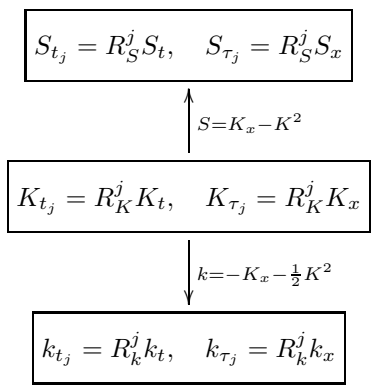
    \mbox{{\bf Diagram 9: The Sawada-Kotera and $k$-equation hierarchies}}\\
%\mbox{{\ }}\\
%A
%%%5\\
\boxed{
\vphantom{\frac{DA}{DB}}
S_{t_j}=R^j_S S_t,\quad S_{\tau_j}=R^j_S S_x
}
%%%5\\
\\
\ar[u]^{\ba{l}
\quad {  }\ea }_{S=K_x-K^2\quad}
%%%\\
%%%%%%%%%%%%%%%%%%%\\
%B
\boxed{
\vphantom{\frac{DA}{DB}}
K_{t_j}=R^j_K K_t,\quad K_{\tau_j}=R^j_K K_x
}
%%%
%%%%%%%%\\
\ar[d]^{
k=-K_x-\frac{1}{2}K^2
}_{ }
%%%%%%%\\
\\
%B
\boxed{
\vphantom{\frac{DA}{DB}}
k_{t_j}=R^j_k k_t,\quad k_{\tau_j}=R^j_k k_x
}
}
\end{displaymath}
%\end{center}
}
}

Below we explicitly list the evolution equations of order 7 which belongs to the hierarchies show in Diagram 9 and Diagram 10,
i.e. the 7th-order equation 
\begin{gather*}
  u_{\tau_1}=R^1_u\,u_x,
\end{gather*}
corresponding to all the 5th-order equations $u_t=F$ in Diagram 6.

\strut\hfill

\noindent
The {\bf 7th-order Kupershmidt equation}, $K_{\tau_1}=R_K\, K_x$, where  $R_K$ is the recursion operator of the Kupershmidt equation
(\ref{kupershmidt}) given in Appendix A, has the following form:
\begin{gather}
  K_{\tau_1}=K_{7x}
  -7K_xK_{5x}
  -7K^2K_{5x}
  -21K_{xx}K_{4x}
  -42KK_{x}K_{4x}
  -14K_{xxx}^2\nn\\[0.3cm]
  \qquad
  -70KK_{xx}K_{xxx}
  -56K_x^2K_{3x}
  +14K^2K_xK_{xxx}
  +14K^4K_{xxx}
  -77K_xK_{xx}^2\nn\\[0.3cm]
  \qquad
  +14K^2K_{xx}^2
  +56KK_x^2K_{xx}
  +112K^3K_xK_{xx}
  +\frac{28}{3}K_x^4
  +84K^2K_x^3\nn\\[0.3cm]
  \qquad
  -\frac{28}{3}K^6K_x.  
\end{gather}  

%\strut\vfill

%\pagebreak

\begin{center}
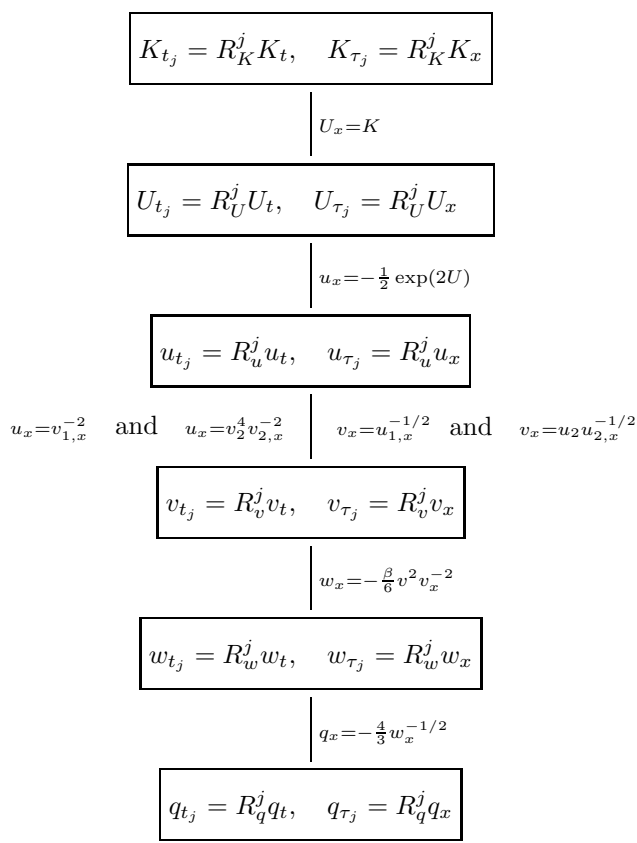
\textcolor{black}{{\bf Diagram 10: A chain of potential Kupershmidt hierarchies}}
\end{center}

%\xymatrix{
%  &Pot. Eq -b\\
%  &\ar@{-}[u]^{a}
%\framebox{Kupershmidt Equation}
%\ar[dr]^{\ \ \ \ \ \ k=-K_x-\frac{1}{2}K^2\ \ \ \ }\ar[dl]_{S=K_x-K^2\ \ \ }\\
%\framebox{Sawada-Kotera Eq}
%&\qquad
%&\framebox{Special Kupershmidt Eq}
%}
%
%\strut\vfill
%
%
%\pagebreak

{\small

  \begin{displaymath}
\qquad\qquad\qquad
    \xymatrix{
&%\hspace{-0.5cm}  
\boxed{
\vphantom{\frac{DA}{DB}}
%\mbox{Kupershmidt Equation: $K$}
K_{t_j}=R^j_KK_t,\quad K_{\tau_j}=R^j_K K_x
}
\\
%%%%%%%%%%%%%%%%%%%%
%  \ar[dr]^{\ \ \ \ \ \ k=-K_x-\frac{1}{2}K^2\ \ \ \ }\ar[dl]_{S=K_x-K^2\ \ \ }\\
%\framebox{Sawada-Kotera Eq: $S$ }
%%%%%%%%%%%%%%%%%%%%
%&\qquad
&\ar@{-}[u]^{\ }_{U_x=K} 
%\boxed{
%\vphantom{\frac{DA}{DB}}
%\mbox{Kupershmidt Eq}}
\boxed{
\vphantom{\frac{DA}{DB}}
%\mbox{1st Potential Eq: $U$}
U_{t_j}=R^j_UU_t,\quad U_{\tau_j}=R^j_U U_x\quad
}
%%%%%%%%%%%%
%&\framebox{$k$-Eq}
& \textcolor{blue}{}  \\
& \ar@{-}[u]_{u_x=-\frac{1}{2}\exp\left(2U\right)}
%\\
%&
%B
\boxed{
\vphantom{\frac{DA}{DB}}
%\mbox{2nd Potential Eq: $u$}
u_{t_j}=R^j_uu_t,\quad u_{\tau_j}=R^j_u u_x
}\\
%%% 1
%%% 1
%A 
      {} &\ar@{-}[u]^{u_x=v_{1,x}^{-2}\ \mbox{\textcolor{black}{{\small \ \ and\ \ }}}\ u_x=v_2^4v_{2,x}^{-2}\ \  }_{\ \ v_x=u_{1,x}^{-1/2}
        \mbox{\textcolor{black}{{\small\ \ and\ \ }}}
        \ v_x=u_2u_{2,x}^{-1/2}}
\boxed{
  \vphantom{\frac{DA}{DB}}
v_{t_j}=R^j_vv_t,\quad v_{\tau_j}=R^j_v v_x  
}
& \textcolor{blue}{}\\
%%%%%%%%%%%%%
%A 
&\ar@{-}[u]_{w_{x}=-\frac{\beta}{6}v^2v_x^{-2}}
\boxed{
\vphantom{\frac{DA}{DB}}
w_{t_j}=R^j_ww_t,\quad w_{\tau_j}=R^j_w w_x  
}
&\textcolor{blue}{}   \\
%A 
&\ar@{-}[u]_{q_x=-\frac{4}{3} w_{x}^{-1/2} }
\boxed{
\vphantom{\frac{DA}{DB}}
q_{t_j}=R^j_qq_t,\quad q_{\tau_j}=R^j_q q_x  
}
& \textcolor{blue}{}
%\\
%C
%A
%%%
%% HERE WE CONTROL THE WIDTH OF THE FIRST COLUMN
%\qquad\qquad\qquad&\ar@{-}[u]_{(w_2)_x=-\frac{16}{9}q_x^{-2} }
%%
%\boxed{
%\vphantom{\frac{DA}{DB}}
%\mbox{$W$-Equation}
%}
%\ar[r]^{w_x=\frac{\beta}{6\lambda_0}\exp\left[-2W\right]\quad\ \ }&\ar@{-}[u]_{w_{2,x}=-\frac{16}{9}q_x^{-2} }
%\boxed{
%\vphantom{\frac{DA}{DB}}
%4^{\mbox{th}}\mbox{-Potential Eq $(-)$: $w_2$}
%}
%& \textcolor{blue}{}
%%% 1
%%% 1
%}
}
\end{displaymath}
}

\strut\hfill

\noindent
The {\bf 7th-order 1st Potential Kupershmidt equation}, $U_{\tau_1}=R_U\, U_x$, where  $R_U$ is the recursion operator of the  1st Potential Kupershmidt equation
(\ref{1st-a-pot-kuper}) given in Appendix A, has the following form:
\begin{gather}
  U_{\tau_1}=U_{7x}
  -7U_{xx}U_{5x}
  -7U_{x}^2U_{5x}
  -14U_{xxx}U_{4x}
  -28U_xU_{xx}U_{4x}
  -21U_xU_{xxx}^2 \nn\\[0.3cm]
  \qquad
-28U_{xx}^2U_{xxx}
  +14U_{x}^2U_{xx}U_{xxx}
  +14U_x^4U_{xxx}
  +\frac{28}{3}U_xU_{xx}^3
  +28U_x^3U_{xx}^2\nn\\[0.3cm]
  \qquad
  -\frac{4}{3}U_x^7.
\end{gather}

\strut\hfill

\noindent
The {\bf 7th-order 2nd Potential Kupershmidt equation}, $u_{\tau_1}=R_u\, u_x$, where  $R_u$ is the recursion operator of the  2nd Potential Kupershmidt equation
(\ref{2nd-pot-kuper}) given in Appendix A, has the following form:
\begin{gather}
  u_{\tau_1}=u_{7x}
  -7u_x^{-1}u_{xx}u_{6x}
  -\frac{35}{2}u_x^{-1}u_{xxx}u_{5x}
  +\frac{147}{4}u_x^{-2}u_{xx}^2u_{5x}
  -\frac{49}{4}u_x^{-1}u_{4x}^2\nn\\[0.3cm]
  \qquad  
  +\frac{301}{2}u_x^{-2}u_{xx}u_{xxx}u_{4x}
  -\frac{609}{4}u_x^{-3}u_{xx}^3u_{4x}
  +\frac{217}{6}u_x^{-2}u_{xxx}^3\nn\\[0.3cm]
  \qquad
  -\frac{1365}{4}u_x^{-3}u_{xx}^2u_{xxx}^2
  +\frac{3675}{8}u_x^{-4}u_{xx}^4u_{xxx}
  -\frac{2457}{16}u_x^{-5}u_{xx}^6.
\end{gather}  

\strut\hfill

\noindent
The {\bf 7th-order 3rd Potential Kupershmidt equation}, $v_{\tau_1}=R_v\, v_x$, where  $R_v$ is the recursion operator of the  3rd Potential Kupershmidt equation
(\ref{3rd-pot-kuper}) given in Appendix A, has the following form:
\begin{gather}
  v_{\tau_1}=v_{7x}
  -7v_x^{-1}v_{xx}v_{6x}
  -7v_x^{-1}v_{xxx}v_{5x}
  +21v_x^{-2}v_{xx}^2v_{5x}
  -7v_x^{-1}v_{4x}^2\nn\\[0.3cm]
  \qquad
  +56v_x^{-2}v_{xx}v_{xxx}v_{4x}
  -42v_x^{-3}v_{xx}^3v_{4x}
  +\frac{14}{3}v_x^{-2}v_{xxx}^3
  -63v_x^{-3}v_{xx}^2v_{xxx}^2\nn\\[0.3cm]
  \qquad
  +42v_x^{-4}v_{xx}^4v_{xxx}.
\end{gather}  

\strut\hfill

\noindent
The {\bf 7th-order 4th Potential Kupershmidt equation}, $w_{\tau_1}=R_w\, w_x$, where  $R_w$ is the recursion operator of the  4th Potential Kupershmidt equation
(\ref{3rd-pot-kuper}) given in Appendix A, has the following form:
\begin{gather}
  w_{\tau_1}=w_{7x}
  -7w_x^{-1}w_{xx}w_{6x}
  -\frac{35}{2}w_x^{-1}w_{xxx}w_{5x}
  +\frac{147}{4}w_x^{-2}w_{xx}^2w_{5x}
  +\frac{7\beta}{6}w_x^{-1}w_{5x}\nn\\[0.3cm]
  \qquad
  -\frac{49}{4}w_x^{-1}w_{4x}^2
  +\frac{301}{2}w_x^{-2}w_{xx}w_{xxx}w_{4x}
  -\frac{609}{4}w_x^{-3}w_{xx}^3w_{4x}
  -\frac{91\beta}{12}w_x^{-2}w_{xx}w_{4x}\nn\\[0.3cm]
  \qquad
   +\frac{217}{6}w_x^{-2}w_{xxx}^3
  -\frac{1365}{4}w_x^{-3}w_{xx}^2w_{xxx}^2
  -\frac{119\beta}{24}w_x^{-2}w_{xxx}^2
  +\frac{3675}{8}w_x^{-4}w_{xx}^4w_{xxx}\nn\\[0.3cm]
  \qquad
  +28\beta w_x^{-3}w_{xx}^2w_{xxx}
  +\frac{7\beta^2}{18}w_x^{-2}w_{xxx}
  -\frac{2457}{16}w_x^{-5}w_{xx}^6
  -\frac{539\beta}{32}w_x^{-4}w_{xx}^4\nn\\[0.3cm]
  \qquad
  -\frac{7\beta^2}{8}w_x^{-3}w_{xx}^2
  -\frac{7\beta^3}{324}w_x^{-2}.
\end{gather}

\strut\hfill

\noindent
The {\bf 7th-order 5th Potential Kupershmidt equation}, $q_{\tau_1}=R_q\, q_x$, where  $R_q$ is the recursion operator of the  5th Potential Kupershmidt equation
(\ref{5th-pot-kuper}) given in Appendix A, has the following form:
\begin{gather}
  q_{\tau_1}=q_{7x}
  -7q_x^{-1}q_{xx}q_{6x}
  +21q_x^{-2}q_{xx}^2q_{5x}
  -7q_x^{-1}q_{xxx}q_{5x}
  +\frac{21\beta}{32}q_{x}^2q_{5x}
  -7q_x^{-1}q_{4x}^2\nn\\[0.3cm]
  \qquad
  +56q_x^{-2}q_{xx}q_{xxx}q_{4x}
  -42q_x^{-3}q_{xx}^3q_{4x}
  -\frac{21\beta}{16}q_{x}q_{xx}q_{4x}
  +\frac{14}{3}q_x^{-2}q_{xxx}^3\nn\\[0.3cm]
  \qquad
  -63q_x^{-3}q_{xx}^2q_{xxx}^2
  +\frac{21\beta}{32}q_xq_{xxx}^2
  +42q_x^{-4}q_{xx}^4q_{xxx}
  +\frac{63\beta^2}{512}q_x^4q_{xxx}\nn\\[0.3cm]
  \qquad
  +\frac{9\beta^3}{8192}q_x^7.
\end{gather}  

\strut\hfill

\noindent
The {\bf 7th-order Sawada-Kotera equation}, $S_{\tau_1}=R_S\, S_x$, where  $R_S$ is the recursion operator of the Sawada-Kotera equation
(\ref{Sawada-Kotera}) given in Appendix A, has the following form:
\begin{gather}
%  \label{SKI-eq}
  S_{\tau_1}=S_{7x}
  +7SS_{5x}
  +14S_xS_{4x}
  +14S^2S_{xxx}
  +21S_{xx}S_{xxx}
  +42SS_xS_{xx}\nn\\[0.3cm]
\label{SKI-eq}
  \qquad
  +7S_x^3
  +\frac{28}{3}S^3S_x.  
\end{gather}  
Note that (\ref{SKI-eq}) appears in the literature \cite{Ito-1980} and has been named the {\bf Sawada-Kotera-Ito equation} (see 
\cite{Goktas-Hereman}, equation (3.10) with condition (3.11)).

\strut\hfill

\noindent
The {\bf 7th-order $k$-equation}, $k_{\tau_1}=R_k\, k_x$, where  $R_k$ is the recursion operator of the $k$-equation
(\ref{private-Kuper}) given in Appendix A, has the following form:
\begin{gather}
%  \label{7th-k-eq}
  k_{\tau_1}=k_{7x}+14kk_{5x}
  +49k_xk_{4x}
  +56k^2k_{xxx}
  +84k_{xx}k_{xxx}
  +252kk_xk_{xx}\nn\\[0.3cm]
\label{7th-k-eq}
  \qquad
  +70k_x^3+\frac{244}{3}k^3k_x.
\end{gather}
Note that (\ref{7th-k-eq}) appears in the literature, namely in \cite{Goktas-Hereman} (see equation (3.10) with condition (3.12)).

\section{Concluding remarks}
In this paper we review some of our earlier results on the multipotentialisation of symmetry-integrable equations, but we also report several new
results on the multipotentialisation of the Kupershmidt equation. In fact, Proposition 4 is new, except for case g) which was reported
earlier in \cite{E-E-Reyes} (where the result was derived in a different manner). As far as we know, the recursion operators listed in Appendix A have
not been reported before,
except for the recursion operators of the Kupershmidt equation (\ref{kupershmidt}) \cite{Sanders-Wang-1998} \cite{ME-NE-Class-RO-2011},
the 1st potential Kupershmidt equation (\ref{1st-a-pot-kuper}) \cite{ME-NE-Class-RO-2011},
and the Sawada-Kotera equation (\ref{Sawada-Kotera}) \cite{FO82}. The 7th-order equations that result as the second member of the Kupershmidt chain of
multipotentialisations are listed explicitly. As far as we know, only two of the 7th-order equations that are listed in Section 4 appeared earlier in the
literature, namely the 7th-order equation of the Sawada-Kotera hierarchy (also known as the Sawada-Kotera-Ito equation) and the 7th-order equation of
the $k$-equation hierarchy. 

It is clear that this multipotentialisation procedure for symmetry-integrable
equations is useful, as it can easily lead to interesting nonlocal invariance relations for all the equations in the chain that can be applied to
generate solutions. This was demonstrated here for the Kupershmidt chain of multipotentialisations. We should however point out that not every
symmetry-integrable equation in $1+1$ dimensions can be potentialised. In Appendix B we exemplify the statement of a non-potentialisable equation
for a special Krichever-Novikov equation. It may therefore be of interest to study the procedure of multipotentialisation for the whole class of
symmetry-integrable equations of dimension $1+1$ and classify the
equations accordingly.

The connections between the multipotentialisation process and nonlocal symmetries also needs to be studied in more detail.
In \cite{E-E-Reyes} we have shown this connection for the Kaup-Kupershmidt equation and for the Sawada-Kotera equation (see also \cite{Reyes-2005} regarding
nonlocal symmetries of the Kaup-Kupershmidt equation). Furthermore, it may be of interest to investigate this multipotentialisation procedure for
non-evolutionary equations, systems of equations, and of course for higher-dimensional equations.

\section*{Appendix A: A list of recursion operators}

In this Appendix we list the recursion operators of the equations in Diagram 6. All equations in Diagram 6 admit
six-order integro-differential recursion operators of the form
\begin{gather}
\label{RO-GEN}
R_u=\sum_{j=0}^6G_jD_x^j+\sum_{i=1}^3I_i(u_x,u_t)D_x^{-1}\circ \Lambda_i,\tag{A.1}
\end{gather}
where $G_j$ are functions of $u$ and $x$-derivatives of $u$ for the equation $u_t=F$. 
%$D_x$ is the total derivative operator, $D_x^{-1}$ the integral
%operator and $\Lambda_i$ are integrating factors of the
%equations. 

\strut\hfill

\noindent
    {\bf The Kupershmidt equation} (\ref{kupershmidt}) in the dependent variable $K$ 
   admits the recursion operator $R_K$ of the form (\ref{RO-GEN}) (\cite{Sanders-Wang-1998}, \cite{ME-NE-Class-RO-2011})
  with
\begin{gather*}
  G_0=
 -K_{5x}
 -12KK_{4x}
 -23K_xK_{xxx}
 +3K^2K_{xxx}
  -15K_{xx}^2
 +38KK_xK_{xx}\\[0.3cm]
\qquad
 +38K^3K_{xx}
 +6K_x^3
 +74K^2K_x^2
 -4K^6\\[0.3cm]
 G_1=
 -6K_{4x}-30KK_{xxx}-63K_xK_{xx}
 +9K^2K_{xx}+18KK_x^2+54K^3K_x\\[0.3cm]
 G_2=
 -14K_{xxx}
 -40KK_{xx}-31K_x^2+6K^2K_x+9K^4\\[0.3cm]
 G_3=-15K_{xx}-30KK_x\\[0.3cm]
 G_4=-6K_x-6K^2,\qquad G_5=0,\qquad G_6=1\\[0.3cm]
 \Lambda_1=-2K_{4x}+10K_xK_{xx}+10K^2K_{xx}
 +10KK_x^2-2K^5\\[0.3cm]
 \Lambda_2=-2K,\quad \Lambda_3=0,\quad I_1=K_x,\quad I_2=K_t.
 \end{gather*}

\noindent
The {\bf 1st Potential Kupershmidt equation} (\ref{1st-a-pot-kuper})
in the variable $U$ 
admits the recursion operator $R_U$ of the form (\ref{RO-GEN}) \cite{ME-NE-Class-RO-2011}
with
\begin{gather*}
  G_0=
  -4U_xU_{5x}
  +20U_xU_{xx}U_{xxx}
  +20U_x^3U_{xxx}
  +20U_x^2U_{xx}^2
  -4U_x^6\\[0.3cm]
G_1=-U_{5x}
  -8U_xU_{4x}
  -15U_{xx}U_{xxx}
  +3U_x^2U_{xxx}
  +6U_xU_{xx}^2
  +18U_x^3U_{xx}\\[0.3cm]
G_2=
  -5U_{4x}
  -22U_xU_{xxx}
  -13U_{xx}^2
  +9U_x^4
  +6U_x^2U_{xx}\\[0.3cm]
G_3=
  -9U_{xxx}
  -18U_xU_{xx},\qquad G_4=-6U_{xx}-6U_x^2,\qquad G_5=0,\qquad G_6=1\\[0.3cm]
  \Lambda_1
  =U_{6x}
  -5U_{xx}U_{4x}
  -5U_x^2U_{4x}
  -5U_{xxx}^2
  -20U_xU_{xx}U_{xxx}\\[0.5cm]
\qquad
-5U_{xx}^3
+5U_x^4U_{xx}\\[0.3cm]
\Lambda_2=U_{xx},\quad
\Lambda_3=0,\quad
I_1=2U_x,\quad
I_2=2U_t.
\end{gather*}

\strut\hfill

\noindent
The {\bf 2nd Potential Kupershmidt equation} (\ref{2nd-pot-kuper})
admits the recursion operator $R_u$ of the form (\ref{RO-GEN})
with
\begin{gather*}
  G_0=\frac{1}{2}u_x^{-1}u_{7x}
  -3u_x^{-2}u_{xx}u_{6x}
  +\frac{1}{8}\left(
  \vphantom{\frac{DA}{DB}}
  121u_x^{-3}u_{xx}^2u_{5x}
  -66u_x^{-2}u_{xxx}u_{5x}\right)\\[0.3cm]
  \qquad
  -\frac{25}{4}u_x^{-2}u_{4x}^2
  -\frac{125}{2}u_x^{-4}u_{xx}^3u_{4x}
  +\frac{135}{2}u_x^{-3}u_{xx}u_{xxx}u_{4x}
  +\frac{65}{4}u_x^{-3}u_{xxx}^3\\[0.3cm]
  \qquad
  -\frac{575}{4}u_x^{-4}u_{xx}^2u_{xxx}^2
  +\frac{2935}{16}u_x^{-5}u_{xx}^4u_{xxx}
  -\frac{945}{16}u_x^{-6}u_{xx}^6\\[0.3cm]
G_1=-\frac{3}{2}u_x^{-1}u_{6x}
  +13u_x^{-2}u_{xx}u_{5x}
  +\frac{1}{8}\left(
  \vphantom{\frac{DA}{DB}}
  246u_x^{-2}u_{xxx}u_{4x}
  -649u_x^{-3}u_{xx}^2u_{4x}\right)\\[0.3cm]
  \qquad
  -\frac{257}{2}u_x^{-3}u_{xx}u_{xxx}^2
  +\frac{1313}{4}u_x^{-4}u_{xx}^3u_{xxx}
  -162u_x^{-5}u_{xx}^5\\[0.3cm]
G_2=-5u_x^{-1}u_{5x}
  +\frac{95}{2}u_x^{-2}u_{xx}u_{4x}
  +\frac{137}{4}u_x^{-2}u_{xxx}^2
  -\frac{895}{4}u_x^{-3}u_{xx}^2u_{xxx}\\[0.3cm]
  \qquad
  +\frac{2457}{16}u_x^{-4}u_{xx}^4\\[0.3cm]
G_3=-\frac{25}{2}u_x^{-1}u_{4x}
  +78u_x^{-2}u_{xx}u_{xxx}
  -\frac{159}{2}u_x^{-3}u_{xx}^3\\[0.3cm]
 G_4=-12u_x^{-1}u_{xxx}
 +\frac{51}{2}u_x^{-2}u_{xx}^2,\qquad
G_5=-6u_x^{-1}u_{xx},\quad  G_6=1\\[0.3cm]
  \Lambda_1=u_x^{-2}u_{8x}
  -8u_x^{-3}u_{xx}u_{7x}
  -\frac{47}{2}u_x^{-3}u_{xxx}u_{6x}
  +\frac{197}{4}u_x^{-4}u_{xx}^2u_{6x}
  -\frac{85}{2}u_x^{-3}u_{4x}u_{5x}\\[0.3cm]
  \qquad
    +255u_x^{-4}u_{xx}u_{xxx}u_{5x}
    -255u_x^{-5}u_{xx}^3u_{5x}
    +\frac{355}{2}u_x^{-4}u_{xx}u_{4x}^2
    +\frac{995}{4}u_x^{-4}u_{xxx}^2u_{4x}\\[0.3cm]
  \qquad
    -1565u_x^{-5}u_{xx}^2u_{xxx}u_{4x}
    +\frac{16665}{16}u_x^{-6}u_{xx}^4u_{4x}
    -760u_x^{-5}u_{xx}u_{xxx}^3\\[0.3cm]
  \qquad
    +\frac{12415}{4}u_x^{-6}u_{xx}^3u_{xxx}^2
    -\frac{12435}{4}u_x^{-7}u_{xx}^5u_{xxx}
    +\frac{14175}{16}u_x^{-8}u_{xx}^7   \\[0.3cm]
    \Lambda_2=u_x^{-2}u_{4x}
    -4u_x^{-3}u_{xx}u_{xxx}+3u_x^{-4}u_{xx}^3,\quad \Lambda_3=0,\quad
     I_1=-\frac{1}{2}u_x,\quad I_2=-\frac{1}{2}u_t
    \end{gather*}

\strut\hfill

\noindent
The {\bf 3rd Potential Kupershmidt equation} (\ref{3rd-pot-kuper})
in the variable $v$ 
admits the recursion operator $R_v$ of the form (\ref{RO-GEN})
with
\begin{gather*}
  G_0=2v_x^{-1}v_{7x}
  -12v_x^{-2}v_{xx}v_{6x}
  +38v_x^{-3}v_{xx}^2v_{5x}
  -18v_x^{-2}v_{xxx}v_{5x}
  -10v_x^{-2}v_{4x}^2\\[0.3cm]
  \qquad
  -70v_x^{-4}v_{xx}^3v_{4x}
  +90v_x^{-3}v_{xx}v_{xxx}v_{4x}
  +20v_x^{-3}v_{xxx}^3
  -110v_x^{-4}v_{xx}^2v_{xxx}^2\\[0.3cm]
  \qquad
  +70v_x^{-5}v_{xx}^4v_{xxx}\\[0.3cm]
  G_1=-3v_x^{-1}v_{6x}+16v_x^{-2}v_{xx}v_{5x}
  +21v_x^{-2}v_{xxx}v_{4x}-41v_x^{-3}v_{xx}^2v_{4x}\\[0.3cm]
  \qquad
  -46v_x^{-3}v_{xx}v_{xxx}^2
  +53v_x^{-4}v_{xx}^3v_{xxx}\\[0.3cm]
  G_2=-2v_x^{-1}v_{5x}+19v_x^{-2}v_{xx}v_{4x}
  +2v_x^{-2}v_{xxx}^2-31v_x^{-3}v_{xx}^2v_{xxx}\\[0.3cm]
 G_3=24v_x^{-2}v_{xx}v_{xxx}-12v_x^{-3}v_{xx}^3-8v_x^{-1}v_{4x}\\[0.3cm]
 G_4=-3v_x^{-1}v_{xxx}+12v_x^{-2}v_{xx}^2,\quad
G_5=-6v_x^{-1}v_{xx},\quad  G_6=1\\[0.3cm]
%%%
  \Lambda_1=v_x^{-2}v_{8x}-8v_x^{-3}v_{xx}v_{7x}
  -16v_x^{-3}v_{xxx}v_{6x}
  +38v_x^{-4}v_{xx}^2v_{6x}
  -20v_x^{-3}v_{4x}v_{5x}\\[0.3cm]
\qquad  +120v_x^{-4}v_{xx}v_{xxx}v_{5x}
-120v_x^{-5}v_{xx}^3v_{5x}
+65v_x^{-4}v_{xx}v_{4x}^2
+80v_x^{-4}v_{xxx}^2v_{4x}\\[0.3cm]
\qquad -440v_x^{-5}v_{xx}^2v_{xxx}v_{4x}
+240 v_x^{-6}v_{xx}^4v_{4x} 
-160v_x^{-5}v_{xx}v_{xxx}^3
+460v_x^{-6}v_{xx}^3v_{xxx}^2\\[0.3cm]
\qquad
-240v_x^{-7}v_{xx}^5v_{xxx}\\[0.3cm]
\Lambda_2=v_x^{-2}v_{4x}-4v_x^{-3}v_{xx}v_{xxx}+3v_x^{-4}v_{xx}^3,\quad \Lambda_3=0,\quad
 I_1=-2v_x,\quad I_2=-2v_t.
\end{gather*}

\strut\hfill

\noindent
The {\bf 4th  Potential Kupershmidt equation} (\ref{4th-pot-kuper}) 
in the variable $w$
admits the recursion operator $R_w$ of the form (\ref{RO-GEN})
with
\begin{gather*}
\hspace{-0.5cm} G_0=\frac{1}{2}w_x^{-1}w_{7x}
  -3w_x^{-2}w_{xx}w_{6x}
  -\frac{33}{4}w_{x}^{-2}w_{xxx}w_{5x}+\frac{121}{8}w_x^{-3}w_{xx}^2w_{5x}
 +\frac{\beta}{3}w_x^{-2}w_{5x}\\[0.3cm]
  \qquad
   +\frac{135}{2}w_x^{-3}w_{xx}w_{xxx}w_{4x}
   -\frac{125}{2}w_x^{-4}w_{xx}^3w_{4x}
   -\frac{13\beta}{6}w_x^{-3}w_{xx}w_{4x}
 -\frac{25}{4}w_x^{-2}w_{4x}^2
   \\[0.3cm]
   \qquad
    +\frac{65}{4}w_x^{-3}w_{xxx}^3
   -\frac{575}{4}w_x^{-4}w_{xx}^2w_{xxx}^2
 -\frac{29\beta}{24}w_x^{-3}w_{xxx}^2
 +\frac{2935}{16}w_x^{-5}w_{xx}^4w_{xxx}\\[0.3cm]
 \qquad
 +\frac{115\beta}{16}w_x^{-4}w_{xx}^2w_{xxx}
 -\frac{5\beta^2}{72}w_x^{-3}w_{xxx}
 -\frac{945}{16}w_x^{-6}w_{xx}^6
 -\frac{25\beta}{6}w_x^{-5}w_{xx}^4\\[0.3cm]
 \qquad
 +\frac{5\beta^2}{144}w_x^{-4}w_{xx}^2
 +\frac{\beta^3}{54}w_x^{-3}\\[0.3cm]
  G_1=-\frac{3}{2}w_{x}^{-1}w_{6x}
  +13w_{x}^{-2}w_{xx}w_{5x}
  +\frac{123}{4}w_x^{-2}w_{xxx}w_{4x}
  -\frac{649}{8}w_x^{-3}w_{xx}^2w_{4x}\\[0.3cm]
  \qquad
  -\frac{9\beta}{4}w_x^{-2}w_{4x}
  -\frac{257}{2}w_x^{-3}w_{xx}w_{xxx}^2
  +\frac{1313}{4}w_x^{-4}w_{xx}^3w_{xxx}
  +\frac{44\beta}{3}w_x^{-3}w_{xx}w_{xxx}\\[0.3cm]
  \qquad
  -162w_x^{-5}w_{xx}^5
  -\frac{49\beta}{3}w_x^{-4}w_{xx}^3
  -\frac{2\beta^2}{3}w_x^{-3}w_{xx}\\[0.3cm]
  G_2=-5w_{x}^{-1}w_{5x}
  +\frac{95}{2}w_x^{-2}w_{xx}w_{4x}
  +\frac{137}{4}w_x^{-2}w_{xxx}^2
  -\frac{895}{4}w_x^{-3}w_{xx}^2w_{xxx}\\[0.3cm]
  \qquad
  -\frac{49\beta}{12}w_x^{-2}w_{xxx}
  +\frac{2457}{16}w_x^{-4}w_{xx}^4
  +\frac{143\beta}{12}w_x^{-3}w_{xx}^2
  +\frac{\beta^2}{4}w_x^{-2}\\[0.3cm]
G_3=-\frac{25}{2}w_x^{-1}w_{4x}
  +78w_x^{-2}w_{xx}w_{xxx}
  -\frac{159}{2}w_x^{-3}w_{xx}^3
  -5\beta w_x^{-2}w_{xx}\\[0.3cm]
G_4=-12w_x^{-1}w_{xxx}+\frac{51}{2}w_x^{-2}w_{xx}^2+\beta w_x^{-1},\quad G_5=-6w_x^{-1}w_{xx},\quad G_6=1\\[0.3cm]
  \Lambda_1=w_x^{-2}w_{8x}
  -8w_x^{-3}w_{xx}w_{7x}
  -\frac{47}{2}w_x^{-3}w_{xxx}w_{6x}
  +\frac{197}{4}w_x^{-4}w_{xx}^2w_{6x}
  +\beta w_x^{-3}w_{6x}\\[0.3cm]
  \qquad
  -\frac{85}{2}w_x^{-3}w_{4x}w_{5x}
  +255 w_x^{-4}w_{xx}w_{xxx}w_{5x}
  -255w_x^{-5}w_{xx}^3w_{5x}
  -9\beta w_x^{-4}w_{xx}w_{5x}\\[0.3cm]
  \qquad
  +\frac{355}{2}w_x^{-4}w_{xx}w_{4x}^2
+\frac{995}{4}w_x^{-4}w_{xxx}^2w_{4x}
-1565w_x^{-5}w_{xx}^2w_{xxx}w_{4x}\\[0.3cm]
  \qquad
-15\beta w_x^{-4}w_{xxx}w_{4x}
+\frac{553\beta}{12}w_x^{-5}w_{xx}^2w_{4x}
+\frac{16665}{16}w_x^{-6}w_{xx}^4w_{4x}
+\frac{5\beta^2}{18}w_x^{-4}w_{4x}\\[0.3cm]
  \qquad
-760 w_x^{-5}w_{xx}w_{xxx}^3
+\frac{12415}{4}w_x^{-6}w_{xx}^3w_{xxx}^2
+\frac{373\beta}{6}w_x^{-5}w_{xx}w_{xxx}^2\\[0.3cm]
  \qquad
-\frac{12435}{4}w_x^{-7}w_{xx}^5w_{xxx}
-\frac{965\beta}{6}w_x^{-6}w_{xx}^3w_{xxx}
-\frac{20\beta^2}{9}w_x^{-5}w_{xx}w_{xxx}\\[0.3cm]
\qquad
+\frac{14175}{16}w_x^{-8}w_{xx}^7
+\frac{605\beta}{8}w_x^{-7}w_{xx}^5
+\frac{25\beta^2}{9}w_x^{-6}w_{xx}^3
+\frac{\beta^3}{27}w_x^{-5}w_{xx}\\[0.3cm]
\Lambda_2=w_x^{-2}w_{4x}
  -4w_x^{-3}w_{xx}w_{xxx}
  +3w_x^{-4}w_{xx}^3
  +\frac{2\beta}{3}w_x^{-3}w_{xx},\quad \Lambda_3=0\\[0.3cm]
  I_1=-\frac{1}{2}w_x,\quad I_2=-\frac{1}{2}w_t.
\end{gather*}

\noindent
The  {\bf 5th Potential Kupershmidt equation} (\ref{5th-pot-kuper})
in the variable $q$
admits the recursion operator $R_q$ of the form (\ref{RO-GEN})
with
\begin{gather*}
  G_0=2q_x^{-1}q_{7x}
  -12q_x^{-2}q_{xx}q_{6x}
  -18q_x^{-2}q_{xxx}q_{5x}
  +38q_x^{-3}q_{xx}^2q_{5x}
  +\frac{21\beta}{16}q_xq_{5x}\\[0.3cm]
  \qquad
  -10q_x^{-2}q_{4x}^2
  +90q_x^{-3}q_{xx}q_{xxx}q_{4x}
  -70q_x^{-4}q_{xx}^3q_{4x}
  -\frac{15\beta}{16}q_{xx}q_{4x}
  +20q_x^{-3}q_{xxx}^3\\[0.3cm]
  \qquad
  -110w_x^{-4}q_{xx}^2q_{xxx}^2
  +\frac{15\beta}{16}q_{xxx}^2
  +70q_x^{-5}q_{xx}^4q_{xxx}
  -\frac{45\beta}{16}q_x^{-1}q_{xx}^2q_{xxx}\\[0.3cm]
  \qquad
  +\frac{9\beta^2}{32}q_x^3q_{xxx}
  +\frac{15\beta}{8}q_x^{-2}q_{xx}^4
  +\frac{9\beta^2}{128}q_x^2q_{xx}^2
  +\frac{27\beta^3}{8192}q_x^6\\[0.3cm]
  G_1=-3q_x^{-1}q_{6x}
  +16q_x^{-2}q_{xx}q_{5x}
  +21q_x^{-2}q_{xxx}q_{4x}
  -41q_x^{-3}q_{xx}^2q_{4x}
  -\frac{27\beta}{32}q_xq_{4x}\\[0.3cm]
\qquad
  -46q_x^{-3}q_{xx}q_{xxx}^2
  +53q_x^{-4}q_{xx}^3q_{xxx}
  -\frac{15\beta}{32}q_{xx}q_{xxx}
  +\frac{15\beta}{16}q_x^{-1}q_{xx}^3\\[0.3cm]
\qquad
  -\frac{27\beta^2}{512}q_x^3q_{xx}\\[0.3cm]
  G_2=-2q_x^{-1}q_{5x}
  +19q_z^{-2}q_{xx}q_{4x}
  +2q_x^{-2}q_{xxx}^2
  -31q_x^{-3}q_{xx}^2q_{xxx}
  +\frac{39\beta}{32}q_xq_{xxx}\\[0.3cm]
  \qquad
  -\frac{15\beta}{32}q_{xx}^2
  +\frac{81\beta^2}{1024}q_x^4\\[0.3cm]
  G_3=-8q_x^{-1}q_{4x}
  +24q_x^{-2}q_{xx}q_{xxx}
  -12q_x^{-3}q_{xx}^3
  -\frac{9\beta}{8}q_xq_{xx}\\[0.3cm]
  G_4=-3q_x^{-1}q_{xxx}
  +12q_x^{-2}q_{xx}^2
  +\frac{9\beta}{16}q_x^2,\quad
  G_5=-6q_x^{-1}q_{xx},\quad G_6=1\\[0.3cm]
  \Lambda_1=q_x^{-2}q_{8x}
  -8q_x^{-3}q_{xx}q_{7x}
  -16q_x^{-3}q_{xxx}q_{6x}
  +38q_x^{-4}q_{xx}^2q_{6x}
  +\frac{9\beta}{16}q_{6x}\\[0.3cm]
  \qquad
  -20q_x^{-3}q_{4x}q_{5x}
  +120q_x^{-4}q_{xx}q_{xxx}q_{5x}
  -120q_x^{-5}q_{xx}^2q_{5x}
  +65q_x^{-4}q_{xx}q_{4x}^2\\[0.3cm]
  \qquad
  +80q_x^{-4}q_{xxx}^2q_{4x}
  -440q_x^{-5}q_{xx}^2q_{xxx}q_{4x}
  +240q_x^{-6}q_{xx}^4q_{4x}
  -\frac{15\beta}{16}q_x^{-2}q_{xx}^2q_{4x}\\[0.3cm]
  \qquad
  +\frac{45\beta^2}{512}q_x^2q_{4x}
  -160q_x^{-5}q_{xx}q_{xxx}^3
  +460q_x^{-6}q_{xx}^3q_{xxx}^2
  -\frac{15\beta}{8}q_x^{-2}q_{xx}q_{xxx}^2\\[0.3cm]
  \qquad
  -240q_x^{-7}q_{xx}^5q_{xxx}
  +\frac{15\beta}{4}q_x^{-3}q_{xx}^3q_{xxx}
  +\frac{45\beta^2}{128}q_xq_{xx}q_{xxx}
  -\frac{45\beta}{32}q_x^{-4}q_{xx}^5\\[0.3cm]
  \qquad
  +\frac{45\beta^2}{512}q_{xx}^3
  +\frac{135\beta^3}{32768}\,q_x^4q_{xx}\\[0.3cm]
\Lambda_2=q_x^{-2}q_{4x}
  -4q_x^{-3}q_{xx}q_{xxx}
  +3q_x^{-4}q_{xx}^3
  +\frac{3\beta}{32}q_{xx}\\
\Lambda_3=0,\quad I_1=-2q_x,\quad  I_2=-2q_t.
\end{gather*}

\strut\hfill

\noindent
The {\bf Sawada-Kotera equation} (\ref{Sawada-Kotera})   in the variable $S$ 
admits the recursion operator $R_S$ of the form (\ref{RO-GEN}) (\cite{SK74}, \cite{CDG76}, \cite{FO82})
with
\begin{gather*}
  G_0=5S_{4x}+16SS_{xx}+6S_x^2+4S^3,\ 
  G_1=10S_{3x}+21SS_x,\ G_2=11S_{xx}+9S^2\\[0.3cm]
 G_3=9S_x,\quad G_4=6S,\quad  G_5=0,\quad G_6=1\\[0.3cm]
 \Lambda_1=2S_{xx}+S^2,\quad \Lambda_2=1,\quad \Lambda_3=0,\quad I_1=S_x,\quad I_2=S_t.
\end{gather*}  

\strut\hfill

\noindent
The {\bf $k$-equation} (\ref{private-Kuper}) in the
variable $k$ 
admits the recursion operator $R_k$ of the form (\ref{RO-GEN})
with
\begin{gather*}
  G_0=13k_{4x}+82kk_{xx}+69k_x^2+32k^3,\quad
  G_1=35k_{xxx}+120kk_x\\[0.3cm]
  G_2=36k^2+49k_{xx},\quad
  G_3=36k_x,\quad G_4=12k,\quad G_5=0,\quad G_6=1\\[0.3cm]
  \Lambda_1=k_{xx}+4k^2,\quad \Lambda_2=1,\quad \Lambda_3=0,\quad I_1=k_x,\quad I_2=k_t.
\end{gather*}

\section*{Appendix B: An equation that does not potentialise}

Not all symmetry-integrable equations can be potentialised. We demonstrate this claim for the following special Krichever-Novikov equation:
\begin{gather}
\label{KN-u3}
u_t=u_{xxx}-\frac{3}{2}\frac{u_{xx}^2}{u_x}+\frac{u^3}{u_x}.    \tag{B.1}
\end{gather}
Equation (\ref{KN-u3}) is a special case of the symmetry-integrable Krichever-Novikov equation \cite{Krichever-Novikov} (see also \cite{M})
\begin{gather}
  \label{KN-gen}
  u_t=u_{xxx}-\frac{3}{2}\frac{u_{xx}^2}{u_x}+\frac{P(u)}{u_x},\qquad P^{(5)}(u)=0. \tag{B.2}
\end{gather}  
In \cite{EE-KN} we report some multipotentialisations of (\ref{KN-gen}), namely for the case
\begin{gather*}
  P(u)=k_2(u^2+k_1u+k_0)^2,
\end{gather*}  
where $k_0,\ k_1$ and $k_2$ are arbitrary constants with $k_2\neq 0$. Equation (\ref{KN-u3}) was not included in our study, as (\ref{KN-u3}) does not admit a
potentialisation, which we'll now establish explicitly.

Calculating all integrating factors for (\ref{KN-u3}) up to order four, we find that (\ref{KN-u3}) admits no zero-order and no second-order integrating factors.
In fact, the lowest integrating factor that (\ref{KN-u3}) admits is a fourth-order integrating factor, namely the following:
\begin{gather*}
  \Lambda=\frac{u_{4x}}{u_x^2}
  -\frac{4u_{xx}u_{3x}}{u_x^3}
  +\frac{3u_{xx}^3}{u_x^4}
  -\frac{2u^3u_{xx}}{u_x^4}
  +\frac{3u^2}{u_x^2}.
\end{gather*}
This gives the following conserved current and flux (with $\Phi^x=\Phi^x(x,u,u_x,\ldots,u_{qx})$, $q=4$ as the lowest order):
\begin{gather*}
  \Phi^t=-\frac{1}{2}\frac{u_{xx}^2}{u_x^2}
  +\frac{1}{3}\frac{u^3}{u_x^2}\\[0.3cm]
  \Phi^x=\frac{u_{4x}u_{xx}}{u_x^2}
  +\frac{1}{2}\frac{u_{xxx}^2}{u_x^2}
  +\frac{2u_{xx}^2u_{xxx}}{u_x^3}
  +\frac{2}{3}\frac{u^3u_{xxx}}{u_x^3}\\[0.3cm]
  \qquad
  -\frac{9}{8}\frac{u_{xx}^4}{u_x^4}
  +\frac{1}{2}\frac{u^3u_{xx}^2}{u_x^4}
  -\frac{3u^2u_{xx}}{u_x^2}
  +\frac{1}{6}\frac{u^6}{u_x^4}.
\end{gather*}
For a potential variable $v$ we need to consider
\begin{gather*}
  v_x=\Phi^t
\end{gather*}
and replace $u$ and the $x$-derivatives of $u$ in
\begin{gather*}
  v_t=-\Phi^x
\end{gather*}
to obain a local equation of the form
\begin{gather*}
  v_t=G(x,v_x,v_{xx},v_{xxx}).
\end{gather*}  
It is now easy to show that this is not possible for the above given $\Phi^t$ and $\Phi^x$.
The same is true for higher-order integrating factors and currents and fluxes that depend on higher-order derivatives.
We therefore conclude that (\ref{KN-u3}) cannot be potentialised.

\section*{Acknowledgement}
It is our pleasure to thank Enrique Reyes for useful comments on a draft version of this paper that led to the current improved version.

\end{document}